\documentclass[12pt,twocolumn,twocolappendix]{aastex631}

\usepackage{amsmath}
\usepackage{hyperref}       
\usepackage{cleveref}
\usepackage{natbib}
\defcitealias{Oden2026}{O26}
\usepackage{float} 
\usepackage[utf8]{inputenc}
\crefformat{MAPS}{\S#2#1#3}

\newcommand{\h}{$^{\rm h}$}
\newcommand{\m}{$^{\rm m}$}
\newcommand{\s}{$^{\rm s}~$}

\usepackage{listings}
\usepackage{xcolor}
\usepackage{animate}

\lstdefinestyle{SQLStyle}{
  language=SQL,
  basicstyle=\ttfamily\footnotesize, 
  keywordstyle=\normalfont,          
  commentstyle=\color{gray},
  stringstyle=\normalfont,
  showstringspaces=false,
  breaklines=true,
  breakindent=1em,
  keepspaces=true
}

\shorttitle{The ringing Large Magellanic Cloud: evidence for a recent dwarf--dwarf collision}

\shortauthors{Oden et al.}

\begin{document}


\title{The ringing Large Magellanic Cloud: evidence for a recent dwarf--dwarf collision}
\author[0009-0000-9037-1697]{Slater J. Oden}
\affiliation{Department of Physics, Montana State University, P.O. Box 173840, Bozeman, MT 59717-3840}
\email{slateroden@montana.edu}

\author[0000-0001-6013-9125]{Bethany Garver}
\affiliation{Department of Physics, Montana State University, P.O. Box 173840, Bozeman, MT 59717-3840}

\author[0000-0002-6553-7082]{Joshua Povick}
\affiliation{Lund Observatory, Division of Astrophysics, Department of Physics, Lund University, Box 118, SE-22100 Lund, Sweden}

\author[0000-0002-1793-3689]{David L. Nidever}
\affiliation{Department of Physics, Montana State University, P.O. Box 173840, Bozeman, MT 59717-3840}

\author[0000-0001-7902-0116]{Victor P. Debattista}
\affiliation{Jeremiah  Horrocks  Institute, University of Lancashire, Preston, PR1 2HE, UK}
\begin{abstract}
    
Galaxy interactions are ubiquitous in the Universe and represent a primary pathway for galaxy growth. Close passages of dwarf galaxies can induce large-scale vertical oscillations in stellar disks, a phenomenon first identified in the Milky Way within the past decade. Here, we use recent red clump distance maps of the Large Magellanic Cloud (LMC) to show that its stellar periphery is vertically offset from the central disk plane by $\sim$1.2 kpc at radii of $\sim$8.5 kpc and exhibits an S-shaped warp reaching $\sim$3.6 kpc at $\sim$11.5 kpc.
We compare these observations to an $N$-body simulation of the LMC--Small Magellanic Cloud (SMC) system, in which a direct collision between the Clouds excites vertical oscillations in the outer LMC disk with amplitudes of $\sim$1~kpc and a period of $\sim$600~Myr. The simulated disk best matches the observed vertical structure $\sim$260~Myr after the collision, although this timing depends on the adopted LMC model. At that time, the simulation reproduces the sign and radial onset of the observed offset and the orientation of the warp, but underpredicts the amplitude of the warp by a factor of $\sim$4--5. These results support a collision-driven origin for the large-scale vertical structure of the LMC disk and demonstrate that dwarf--dwarf interactions can reshape stellar disks on galactic scales.

\end{abstract}



\section{Introduction}
\label{sec:intro}


Galaxy interactions play a central role in galaxy evolution, driving structural transformation, stellar mass growth, and the redistribution of angular momentum throughout cosmic time \citep{Toomre1972,Barnes1992,Schweizer2000}. Although mergers and tidal encounters are common, their detailed dynamical impact on stellar disks is rarely resolved directly, especially in low-mass dwarf galaxies. This limitation is particularly important because dwarf galaxies are the most numerous galaxies in the Universe and are thought to represent the fundamental building blocks of hierarchical galaxy growth \citep{Huvsko2022,Amorisco2014}. Despite the expected frequency of dwarf--dwarf interactions, their ability to reshape stellar disks through tidal perturbations and direct collisions remains poorly constrained observationally.


Numerical simulations predict that interactions with massive satellites can excite large-scale vertical oscillations in galactic disks, producing warps, bending waves, and outer-disk substructure that persist for hundreds of millions of years \citep{Gomez2016,Gomez2017,Laporte2018,McMillan2022}. The Milky Way (MW) provides one of the clearest observational laboratories for studying such vertical disk structure, exhibiting coherent north--south number-count asymmetries and bulk vertical motions \citep{Widrow2012,Yanny2013}, outer-disk corrugations \citep{Xu2015}, and the phase-space spiral revealed by \textit{Gaia} \citep{Antoja2018}. External perturbations from the Sagittarius dwarf galaxy and the Large Magellanic Cloud (LMC) have been proposed as possible drivers of some of these features, including the phase-space spiral, Monoceros Ring, and Triangulum--Andromeda clouds \citep{Momany2006,Gomez2013,Gomez2016,Laporte2018,Bland-Hawthorn2021}. However, their origins remain debated, and internal or secular mechanisms—including bar buckling and bending waves driven by irregular gas inflow along the Galactic warp—may also contribute to or produce similar structures \citep{Wang2026}. Interpreting the MW disk response is further complicated by our location within the disk, dust obscuration, and the lack of an external viewing perspective. Nearby external galaxies therefore provide an important opportunity to map the global three-dimensional response of a stellar disk and, where complementary observations constrain the interaction history, connect that response to a likely external perturbation.


The LMC and Small Magellanic Cloud (SMC) are the nearest interacting massive dwarf galaxies, at heliocentric distances of $\sim$50 and $\sim$60~kpc \citep[hereafter O26]{Oden2026}, and provide a unique laboratory for studying dwarf--dwarf encounters. Their proximity allows individual stars to be resolved across both systems, enabling the three-dimensional structure and kinematics of the Magellanic System to be mapped in exceptional detail. A growing body of work indicates that the Magellanic Clouds (MCs) experienced a recent close passage, possibly a direct collision with an impact parameter of a few kpc, approximately 100--250~Myr ago \citep{Besla2016,Zivick2018,Zivick2019,Choi2022,Dhanush2024}. This interaction history is constrained by complementary lines of evidence, including proper motion (PM) fields along the Magellanic Bridge, disturbed kinematics and line-of-sight structure in the eastern SMC, and young stars and clusters in the Bridge region whose ages are consistent with formation during the most recent encounter \citep{Zivick2018,Zivick2019,Harris2009,JacyszynDobrzeniecka2016,Bitsakis2018}. Such an encounter is expected to perturb the LMC disk, leaving imprints on its stellar kinematics and morphology, including large-scale asymmetries, disturbed outer-disk substructures, and vertical oscillations \citep{Besla2012,Besla2016,Choi2018a,Choi2018b,JimenezArranz2024a}.

The LMC itself has long been known to deviate from a simple, planar disk. It is a barred Magellanic spiral with an off-center bar and a dominant single spiral arm \citep{devaucouleurs1972,Zhao2000,vandermarel2001a}, and its disk exhibits warping, twisting, and radial variations in inclination and line-of-nodes position angle (\citealt{vandermarel2002,Olsen2002,Choi2018a,Saroon2022,JimenezArranz2025}; \citetalias{Oden2026}). Beyond the nominal disk ($\gtrsim$7$^{\circ}$), wide-field surveys have revealed low-surface-brightness features, including the Northern Arm and southern hook-like structures, that have been interpreted as perturbed outer-disk material \citep{Mackey2016,Mackey2017,Belokurov2019,elyoussoufi2021,Cullinane2022a}. Building on these studies, \citetalias{Oden2026} used $\sim$2.3~million red clump (RC) stars from \textit{Gaia} DR3 to construct a contiguous three-dimensional distance map of the Magellanic System, demonstrating that previously identified localized warp features form part of a larger vertical distortion of the LMC disk. However, a direct dynamical connection between this global vertical structure and the recent LMC--SMC interaction has not yet been established.

Here, we investigate whether the recent LMC--SMC collision can explain the large-scale vertical structure of the LMC disk. We build on the methodology of \citetalias{Oden2026}, comparing RC distance maps of the LMC derived from \textit{Gaia} DR3 data (\citealt{GaiaCollaboration2016}; \citetalias{Oden2026}) to the best-fitting model from a published suite of $\sim$8,000 self-consistent $N$-body simulations of the coupled MW--LMC--SMC system spanning the past 2.5~Gyr \citep{Garver2026}. We reconstruct the LMC disk in three dimensions, quantify its vertical displacement field using low-order azimuthal Fourier modes, and compare the resulting mode structure to the simulated response following the recent LMC--SMC collision. We show that the observed disk exhibits an axisymmetric vertical offset at intermediate radii and an integral-sign ($m=1$) warp in the outer disk, and that the simulation reproduces the sign and radial onset of the offset and the orientation of the warp, but not the amplitude of the warp.

This paper is organized
as follows. Section~\ref{sec:data} describes the \textit{Gaia} DR3 red clump sample;
Section~\ref{sec:structure} presents the three-dimensional disk reconstruction and
observed warp morphology; Section~\ref{sec:sims} describes the $N$-body simulation;
Section~\ref{sec:fourier_comp} presents the Fourier decomposition and the
observation--simulation comparison; and Sections~\ref{sec:discussion} and
\ref{sec:conclusion} discuss the results and summarize our conclusions.

\section{Data}
\label{sec:data}

The observational basis of this work is the \textit{Gaia} DR3 RC distance map of the MCs presented in \citetalias{Oden2026}. RC stars provide a powerful tracer of three-dimensional structure because they are numerous, relatively bright, and have a narrow luminosity function \citep{Girardi1999,Girardi2016}. These properties make them especially useful for mapping the low-surface-brightness outskirts of the LMC, where sparse stellar sampling limits many other distance tracers. In this section, we summarize the construction of the LMC RC sample and distance estimates used throughout this work; full details of the selection, extinction correction, population modeling, and validation are given in \citetalias{Oden2026}.

\subsection{Gaia DR3 sample selection}
\label{sec:sample_selection}

The parent catalog is drawn from \textit{Gaia} DR3 \citep{GaiaCollaboration2016,GaiaCollaboration2021a}. We follow the RC selection and distance-mapping methodology of \citetalias{Oden2026}, beginning with sources selected to isolate the Magellanic RC population while minimizing contamination from foreground MW stars and non-RC populations. The initial photometric selection uses magnitude and color cuts of
$17 < G < 20.25$ and $0.5 < (BP-RP) < 2.0$ which encompass the expected location of the Magellanic RC in the \textit{Gaia} color--magnitude diagram (CMD).

\begin{figure*}
\begin{center}
\includegraphics[width=1.0\textwidth]{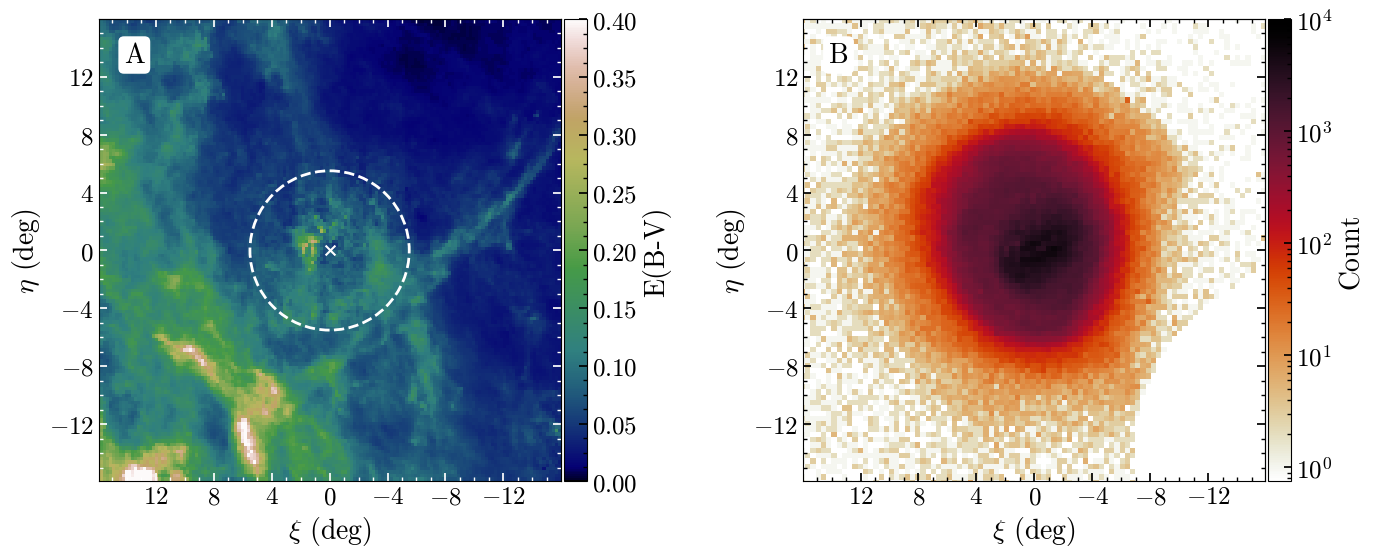}
\caption{Reddening and projected stellar density maps of the LMC red clump sample.
(A) Reddening map $E(B-V)$ across the LMC footprint derived in \citetalias{Oden2026},
shown in the tangent-plane coordinate system $(\xi,\eta)$.
The dashed white circle marks the inner $5.5^{\circ}$ region
centered on the LMC optical center (white cross).
(B) Log-scaled stellar counts per bin of the final RC sample,
illustrating the decline in stellar density from the inner disk
to the outer periphery and motivating the hybrid binning scheme.
We exclude the SMC using a circular cut of radius $13^{\circ}$
in Magellanic Stream coordinates centered on
$(L_{\rm MS},B_{\rm MS})=(-19^{\circ},-12.5^{\circ})$,
chosen to bisect the region of lowest stellar number density
between the Clouds.}
\label{fig:Redding_Density_Maps}
\end{center}
\end{figure*}

Standard astrometric quality cuts are applied to remove sources with problematic astrometric solutions \citep{Lindegren2018}. We require a \texttt{re-normalized unit weight error} (\texttt{ruwe}) $\leq 1.4$, where \texttt{ruwe} is the square root of the reduced $\chi^2$ of the single-star astrometric fit, rescaled to remove its dependence on magnitude and color, so that well-behaved sources have values near 1. We also require \texttt{astrometric\_excess\_noise} $\leq 2$~mas, where this quantity is the additional noise that must be added in quadrature to the astrometric uncertainties of a source for the single-star model to fit its observations. Sources with null parallaxes are excluded, and a parallax signal-to-noise cut of $\varpi/\sigma_{\varpi} < 4$ is imposed to suppress nearby MW foreground stars while retaining Magellanic members with intrinsically low-significance parallaxes. The sample is then refined using iterative PM selections designed to isolate stars with Magellanic-like kinematics, yielding an initial MC RC candidate catalog of $\sim$7.8 million stars. We remove the SMC using a $13^{\circ}$ circular cut in Magellanic Stream coordinates centered on $(L_{\rm MS}, B_{\rm MS}) = (-19^{\circ}, -12.5^{\circ})$, chosen to bisect the region of lowest stellar number density between the Clouds. This spatial separation leaves $\sim$6.6 million LMC RC candidates before the final local CMD-based RC selection.

\subsection{Extinction correction}
\label{sec:extinction_correction}

We adopt the extinction correction framework developed by \citetalias{Oden2026}, which combines multiple reddening maps to account for spatially varying dust across the LMC. In the inner LMC, where high-resolution reddening information is available, the correction uses the OGLE-IV reddening map of \citet{Skowron2021}. In the lower-density outer regions, where OGLE-IV coverage is incomplete, the correction transitions to the all-sky dust map of \citet{Schlegel1998}, recalibrated following \citet{Schlafly2011} to correct the known overestimate in color excess $E(B-V)$.

The OGLE-IV and recalibrated \citet{Schlegel1998} maps are placed on a common $E(B-V)$ scale using the transformations described in \citetalias{Oden2026}. Extinction coefficients are computed assuming the \citet{Fitzpatrick1999} extinction law with $R_V=3.1$, and a smooth logistic weighting function is used to transition between the inner and outer reddening maps. The resulting reddening map is shown in Figure~\ref{fig:Redding_Density_Maps}A. The reddening map is publicly available via Zenodo\footnote{\url{https://doi.org/10.5281/zenodo.18330362}}, with associated code available on GitHub\footnote{\url{https://github.com/slateroden/XMCreddeningmap}} (MIT License).

The inner LMC contains active star-forming regions with elevated and spatially variable extinction that may not be fully captured by the blended map, particularly where the effective extinction law deviates from the assumed $R_V = 3.1$. These systematics introduce biases in the RC distance estimates and the derived disk-frame $Z$ values. We therefore exclude the bar region ($R \lesssim 2$~kpc) from the morphological analysis.

\begin{figure*}
\begin{center}
\includegraphics[width=1.0\textwidth]{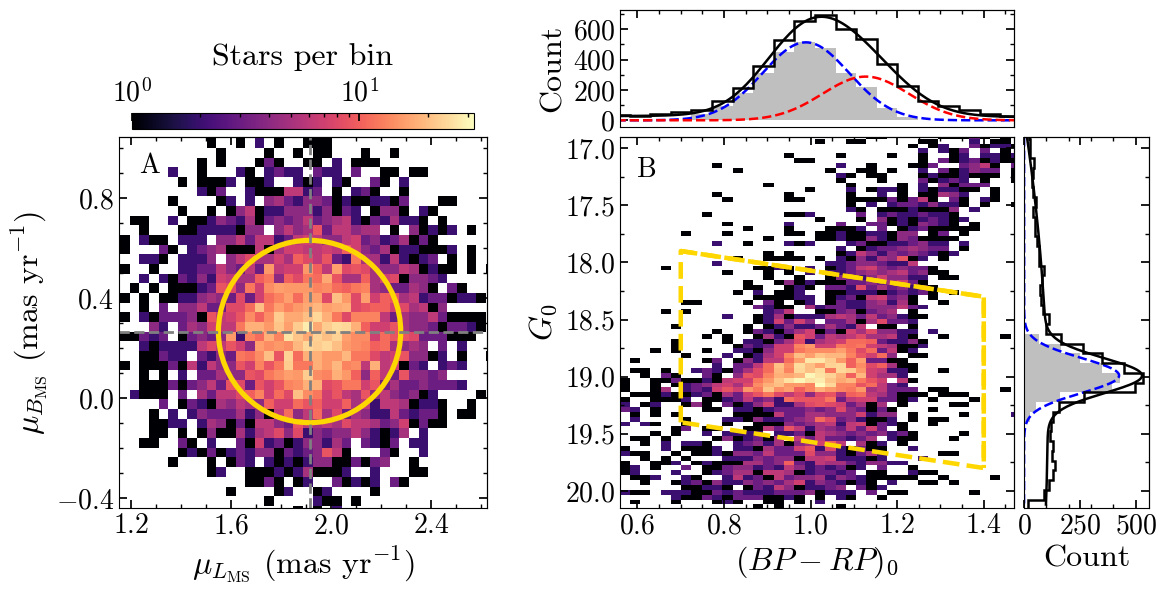}
 \caption{Red clump selection procedures applied to an example field in the outer LMC periphery, centered on $(L_{\rm MS}, B_{\rm MS}) = (-7^{\circ}, 0^{\circ})$ with an aperture radius of $1.5^{\circ}$. The color scale (colorbar above panel A) is shared between the panels and indicates the number of stars per bin on a logarithmic scale.
    (A) Proper motion (PM) diagram in the Magellanic Stream coordinate frame ($\mu_{L_{\rm MS}}$, $\mu_{B_{\rm MS}}$). The yellow circle marks the local kinematic selection aperture, set to twice the robust PM dispersion, used to isolate stars with Magellanic-like kinematics.   The grey dashed lines mark the center of this PM selection.
    (B) Color--magnitude diagram (CMD), showing extinction-corrected magnitude $G_0$ versus $(BP-RP)_0$ color. The yellow dashed box marks the initial CMD selection window used to isolate the red clump population. Marginal histograms (black steps) show the color (top) and magnitude (right) distributions. The solid black curves show the full best-fitting models: a double Gaussian plus a second-order polynomial in color, and a single Gaussian plus a second-order polynomial in magnitude. The blue dashed curves trace the red clump Gaussian component of each fit, and the red dashed curve traces the second Gaussian component in color, corresponding to the red giant branch. The grey shaded region indicates the final $3\sigma$ red clump selection in each marginal distribution.}
\label{fig:cuts}
\end{center}
\end{figure*}

\subsection{Red clump selection and distance calibration}
\label{sec:rcselection_and_distance}

We adopt the spatially adaptive RC selection methodology developed in \citetalias{Oden2026}, which accounts for spatial variations in stellar population properties, line-of-sight distance, and kinematics across the LMC. The intrinsic location of the RC in the CMD varies across the disk due to gradients in age and metallicity, while the LMC’s internal rotation introduces strong spatial structure in the kinematic field. These effects necessitate a locally adaptive selection procedure.

Following \citetalias{Oden2026} we characterize the local RC population on a 0.3$^{\circ}$ grid across the LMC. For each grid point, the stellar sample is selected with an adaptive circular aperture rather than a fixed-area bin. The aperture is initialized at 0.3$^{\circ}$, but in regions where the stellar density is low, it is enlarged in 0.1$^{\circ}$ increments until at least 500 stars are included, with a maximum allowed radius of 5.0$^{\circ}$. This ensures that the dense inner disk is sampled at high angular resolution while the outer periphery retains enough stars for stable CMD and PM characterization. The projected density of the final RC sample, which illustrates the need for this adaptive selection, is shown in Figure~\ref{fig:Redding_Density_Maps}B. PM cuts are then defined locally from each aperture-selected sample, using circular selection regions with radii equal to twice the robust PM dispersion. These cuts reduce MW foreground contamination while retaining $\geq$95\% of Magellanic members \citepalias{Oden2026}. Figure~\ref{fig:cuts}A shows an example of the resulting PM selection for a representative field in the LMC periphery.

\begin{figure*}
\begin{center}
\includegraphics[width=1.0\textwidth]{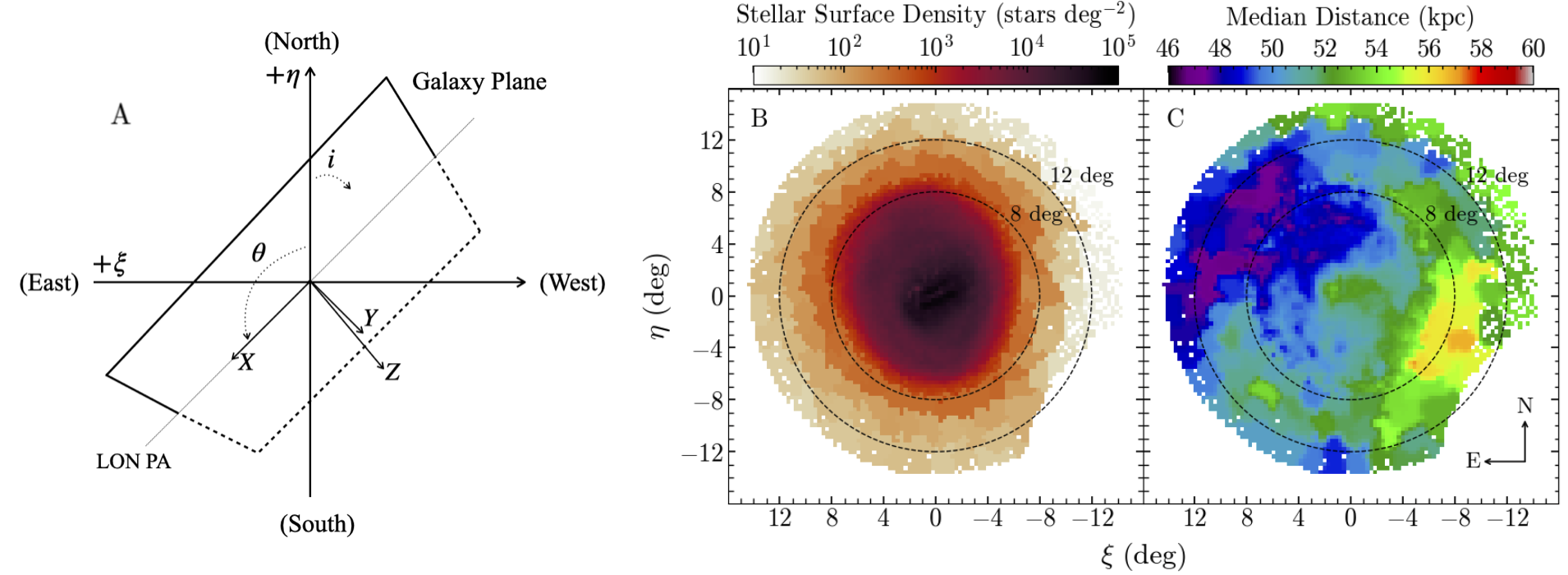}
\caption{Coordinate systems, spatial binning, and heliocentric distance map of the LMC.
    (A) Schematic illustrating the sky-plane tangent coordinates $(\xi,\eta)$ and the disk-plane Cartesian coordinate system $(X,Y,Z)$. The inclination angle $i$ and line-of-nodes position angle $\theta$ define the two successive rotations used to transform into the disk frame, in which $Z$ measures displacement perpendicular to the best-fit disk plane.
    (B) Hybrid binning map in tangent-plane coordinates $(\xi,\eta)$ out to a radius of $14^{\circ}$, colored by stellar surface density (stars per deg$^2$). The scheme transitions from fixed $0.3^{\circ}\times0.3^{\circ}$ grid cells in the dense inner disk to adaptive Voronoi bins in the sparse outer periphery. Dashed circles mark radii of $8^{\circ}$ and $12^{\circ}$.
    (C) Median heliocentric distance of RC stars in each hybrid bin, tracing the three-dimensional structure of the LMC stellar disk.}
\label{fig:Hybrid_binning}
\end{center}
\end{figure*}

Within each aperture-selected sample, the local magnitude and color distributions are modeled using single- and double-Gaussian components, respectively, combined with second-order polynomial backgrounds. The single-Gaussian magnitude fit isolates the RC peak from the underlying red giant branch continuum, while the double-Gaussian color fit separates the RC from the red giant branch. Stars are retained within three standard deviations of the best-fitting Gaussian in both magnitude and color, as illustrated in Figure~\ref{fig:cuts}B. This procedure yields a final sample of $\sim$1.8 million LMC RC stars.

Distances are derived following the RC population-modeling framework of \citetalias{Oden2026}. The intrinsic RC luminosity depends on age and metallicity, so adopting a single fixed absolute magnitude would introduce spatially coherent distance biases. To reduce these effects, \citetalias{Oden2026} first calibrates RC absolute magnitude as a function of intrinsic \textit{Gaia} color using a clean MW RC sample with precise parallaxes \citep{Ruiz-Dern2018}. These initial distances are then refined using spatially resolved star-formation histories derived from deep DECam photometry from the DELVE-MC survey \citep{Nidever2026a}, allowing localized corrections for age-dependent RC luminosity variations.

The resulting distance catalog provides the median RC heliocentric distance across the LMC with sufficient spatial coverage to trace the disk into the low-density periphery. In the following sections, we use this catalog to reconstruct the LMC disk in three dimensions and to quantify its large-scale vertical displacement relative to the inner disk plane.

\section{Three-Dimensional Structure of the LMC Disk}
\label{sec:structure}

We now use the RC distance catalog described in Section~\ref{sec:data} to reconstruct the three-dimensional structure of the LMC disk. In this section, we define the disk-frame coordinate system, describe the tangent-plane hybrid binning scheme used to measure robust median distances, and present the observed large-scale distance structure of the LMC.

\subsection{Disk-frame coordinate transformation}
\label{sec:coordinate_transformation}

To quantify the vertical structure of the LMC, we transform the observed RC distances into an LMC-centered disk-frame coordinate system. Each star is described by its sky position $(\alpha,\delta)$ and heliocentric distance $D$. Following \citet{vandermarel2002} and \citet{JimenezArranz2023}, these quantities are first converted into an LMC-centered on-sky Cartesian frame and then rotated into the disk plane using the line-of-nodes position angle $\theta$ and inclination angle $i$. In the resulting coordinate system, $(X,Y)$ defines the reference disk plane and $Z$ measures vertical displacement above or below that plane, with positive $Z$ corresponding to material displaced toward the observer. We define the disk-frame radius as $R = \sqrt{X^2+Y^2}$ and the disk-plane azimuth $\phi_{\rm disk}$ such that $\phi_{\rm disk}=0^{\circ}$ lies along the negative $Y$ axis and increases counterclockwise in the disk frame. This geometry is illustrated schematically in Figure~\ref{fig:Hybrid_binning}A.

We adopt an LMC center of $(\alpha_0, \delta_0) = (81.28^{\circ}, -69.78^{\circ})$ \citep{vandermarel2001a} and a fiducial distance of $D_0 = 50.1$~kpc. This distance is chosen such that the median vertical displacement of the inner disk is approximately zero, and is consistent with commonly adopted LMC distances from independent tracers, including AGB stars, the tip of the red giant branch, RR Lyrae stars, Cepheids, and eclipsing binaries \citep[e.g.,][]{vandermarel2001a,Borissova2009,Haschke2012a,Ripepi2015,Groenewegen2019,Pietrzynski2019,Graczyk2020,Madore2025}. The disk orientation parameters are determined from an ordinary least-squares plane fit to the RC sample over the radial range $R=2.5$--$4.0$~kpc, where the disk is expected to be approximately planar and minimally affected by the central bar and outer-disk perturbations. This procedure gives $(i,\theta) = (27.52^{\circ},  141.55^{\circ})$.

We verified that our results are insensitive to the choice of LMC center, with commonly used alternatives \citep{Choi2022,Rathore2025} producing consistent disk geometry within our uncertainties. These coordinates provide the basis for the radial and azimuthal analysis of the LMC warp in the following sections.

\subsection{Tangent-plane hybrid binning and line-of-sight distance bias}
\label{sec:hybrid_binning}

A key systematic arises when the disk frame transformation is applied to individual stars. When each star’s distance is rotated into the disk frame on a per-star basis, line-of-sight distance errors propagate into large spurious $|Z|$ values that correlate with on-sky position, producing an artificial vertical signal that can mimic a genuine warp. 

We demonstrate this effect with a controlled test described in detail in Appendix~\ref{app:A} and shown in Figure~\ref{fig:Z_bias_test}. In brief, stars are placed in an infinitesimally thin disk with intrinsic $Z=0$ using their observed on-sky positions and the adopted LMC disk geometry. Gaussian line-of-sight distance scatter is then added to the synthetic distances. Because the input disk has no intrinsic vertical structure, any coherent pattern recovered in $Z$ after transformation must be artificial. Applying the disk-frame transformation on a per-star basis produces a coherent median-$Z$ gradient with amplitudes of several kpc. Binning in distance-dependent Cartesian coordinates $(x,y)$ reduces but does not eliminate the bias, whereas binning in tangent-plane coordinates $(\xi,\eta)$ suppresses the spurious signal and recovers median values consistent with $Z\simeq0$ across the disk.

Motivated by this test, we aggregate stars in tangent-plane coordinates $(\xi,\eta)$ before transforming them into the disk frame. These coordinates depend only on angular position on the sky and are therefore independent of the measured heliocentric distance. For each spatial bin, we compute the median RC distance and then transform the bin into the disk frame using its median sky position. This procedure avoids injecting line-of-sight distance scatter into the spatial coordinates used for the binning itself.

For the main analysis, we use a hybrid binning scheme in $(\xi,\eta)$, shown in Figure~\ref{fig:Hybrid_binning}B. The scheme uses fixed $0.3^{\circ}\times0.3^{\circ}$ bins in the dense inner LMC and transitions to adaptive Voronoi bins in the low-density outer periphery, retaining a minimum of 30 stars per bin. This preserves high spatial resolution in the inner disk while maintaining robust distance estimates at large radii, where the RC sample becomes sparse. We use this tangent-plane hybrid binning scheme for all quantitative measurements of the LMC vertical structure.

\subsection{Observed distance structure}
\label{sec:obsered_distance_structure}

The median RC heliocentric distance map derived from the tangent-plane hybrid binning scheme is shown in Figure~\ref{fig:Hybrid_binning}C. The inner LMC is broadly consistent with an inclined stellar disk, while the outer regions show coherent departures from a simple planar geometry. In particular, the northern and southern outskirts exhibit substantial distance offsets relative to the inner disk, indicating that the LMC periphery is warped and asymmetric.

These large-scale departures become increasingly prominent beyond radii of $\sim$8$^{\circ}$. The outer disk no longer resembles a simple inclined plane, but instead shows coherent distance structure extending around much of the LMC periphery. In the following sections, we transform this binned distance map into disk-frame vertical displacement and compare the resulting morphology to predictions from $N$-body simulation of the recent LMC--SMC interaction.

\begin{figure*}
\begin{center}
\includegraphics[width=1.0\textwidth]{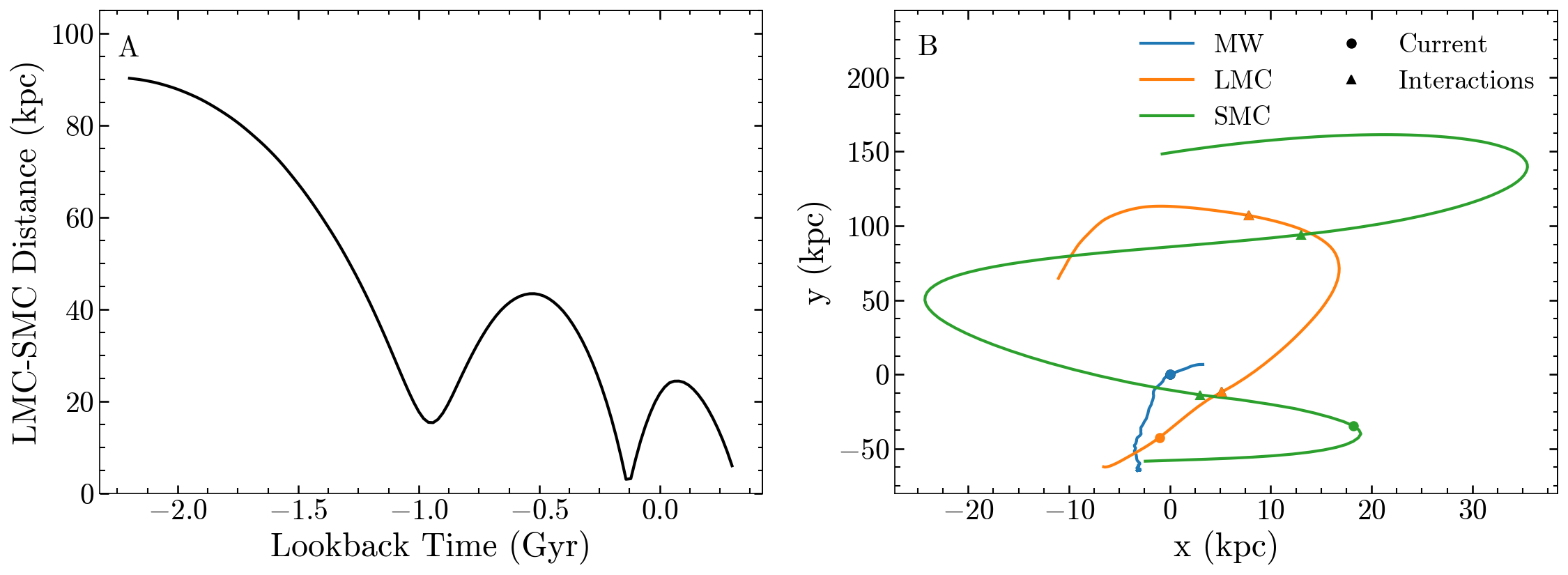}
\caption{LMC--SMC separation and orbital history in the best-fitting orbital model of \citet{Garver2026}, selected to match the present-day positions and velocities of the Clouds (Section~\ref{sec:sims_suite}).
    (A) LMC--SMC separation as a function of lookback time from the nominal present day of the simulation (Section~\ref{sec:sims_orbit}), showing two close passages approximately 940 and 140~Myr earlier. The most recent encounter reaches a minimum separation of $\sim$3~kpc, consistent with a direct collision.
    (B) Projected trajectories of the MW (blue), LMC (orange), and SMC (green) over the past 2.5~Gyr in the MW-centric frame. Filled circles mark the positions of each galaxy at the nominal present day, and filled triangles mark the positions at the times of the two close LMC--SMC passages.}
\label{fig:simDistOrbit}
\end{center}
\end{figure*}

\section{Dynamical simulations of the Magellanic Clouds}
\label{sec:sims}

To interpret the observed three-dimensional structure of the LMC, we compare the observed morphology to a single $N$-body simulation of the Magellanic System drawn from the suite presented by \citet{Garver2026}. The suite models the coupled dynamical evolution of the MW, LMC, and SMC over the past 2.5~Gyr and comprises $\sim$8,000 realizations spanning the orbital parameter space of the Clouds. We use the fiducial high-resolution realization of the orbit that best reproduces the present-day positions and velocities of the Clouds (Section~\ref{sec:sims_suite}), which provides a framework for testing whether the recent LMC--SMC encounter can excite vertical structure comparable to that observed in the LMC disk. Unless otherwise stated, ``the simulation'' refers to this model throughout. Full details are given in \citet{Garver2026}.

\subsection{Simulation suite and galaxy models}
\label{sec:sims_suite}

The simulations were run using the parallel tree code \texttt{pkdgrav} \citep{Stadel2001}. The MW, LMC, and SMC are modeled as live stellar and dark matter systems, allowing tidal deformation, mass loss, and dynamical friction to develop during close passages. This is essential for modeling the response of the LMC disk to the SMC, since fixed-potential orbit integrations cannot capture the coupled structural evolution of the Clouds during a direct encounter.

The simulation suite includes a large set of low-resolution simulations designed to explore the orbital parameter space of the Magellanic System. In these orbital-search simulations, the LMC is sampled with $1.62\times10^{5}$ stellar and $2.38\times10^{5}$ dark matter particles, and the SMC with $2.25\times10^{4}$ stellar and $2.9\times10^{4}$ dark matter particles. All of the orbital-search simulations use the same galaxy masses: the LMC has a stellar mass of $7.20\times10^{9}~M_{\odot}$ embedded in a $1.76\times10^{11}~M_{\odot}$ dark matter halo, and the SMC has a stellar mass of $1.06\times10^{9}~M_{\odot}$ in a $2.02\times10^{10}~M_{\odot}$ dark matter halo. The MW is modeled as a live system with a stellar mass of $5.25\times10^{10}~M_{\odot}$ and a dark matter halo mass of $1.16\times10^{12}~M_{\odot}$.

Each galaxy is first evolved in isolation for 6~Gyr to establish equilibrium. The combined MW--LMC--SMC system is then evolved forward for an additional 2.5~Gyr. Initial conditions are determined using a genetic algorithm that explores a twelve-dimensional parameter space of LMC and SMC initial phase-space coordinates. The full search comprises $\sim$8,000 low-resolution realizations, with the fitness evaluated using only the observed present-day bulk positions and velocities of the Clouds. No constraint on the resulting LMC stellar morphology is imposed during the orbit selection. The simulation therefore provides a physically motivated comparison: the vertical structure of the simulated LMC disk arises as a consequence of an orbit constrained only by the present-day bulk positions and velocities of the Clouds. The best-fitting orbit was then re-simulated with 10 times more particles for a fiducial LMC model (their model A) and for five alternative LMC models that vary the disk thickness, disk velocity dispersion, and dark matter halo concentration. 

In this paper, we analyze the fiducial high-resolution simulation. We use this model because it is the fiducial model of \citet{Garver2026}, on which their comparison with the observations is based. The alternative LMC models were run from the same initial positions and velocities without refitting the orbit, and in several of them the LMC reaches the nominal present day at a different time and the SMC lies at a different position relative to the LMC. We did not choose among the LMC models on the basis of their vertical structure, because that is the quantity we compare with the observations. The dependence of the vertical structure on the LMC model is described in Section~\ref{sec:sims_morphology}.

\subsection{Orbital history}
\label{sec:sims_orbit}

Following \citet{Garver2026}, we define the nominal ``present day'' of the simulation as the epoch at which the simulated LMC crosses $L_{\rm MS}=0^{\circ}$, where $L_{\rm MS}$ is Magellanic Stream longitude \citep{Nidever2008}. At this epoch, the best-fitting orbit places the LMC within 2.8~kpc and 14.8~km~s$^{-1}$ of its observed position and velocity, and the SMC within 13.6~kpc and 13.2~km~s$^{-1}$ \citep{Garver2026}.

The corresponding LMC--SMC separation history is shown in Figure~\ref{fig:simDistOrbit}A. The model contains two close passages between the Clouds, at lookback times of $\sim$940 and $\sim$140~Myr from the nominal present day. The recent passage reaches a minimum LMC--SMC separation of $\sim$3~kpc, consistent with a direct collision. We measured the geometry of this collision from the simulation: at closest approach, the SMC trajectory is inclined by $\sim$83$^{\circ}$ relative to the normal of the LMC disk, corresponding to a nearly coplanar passage only $\sim$7$^{\circ}$ from the disk plane. The corresponding MW-centric orbits of the MW, LMC, and SMC are shown in Figure~\ref{fig:simDistOrbit}B. In the following section, we examine the evolution of the simulated LMC disk following this encounter and identify the evolutionary phase most directly comparable to the observed morphology.


\subsection{Vertical response and comparison snapshot}
\label{sec:sims_morphology}

The recent LMC--SMC collision excites large-scale vertical perturbations in the simulated LMC disk. Following the encounter, the outer disk undergoes a coherent vertical oscillation with a period of $\sim$600~Myr. The response contains both a broad, approximately axisymmetric vertical displacement and a weaker asymmetric warp, whose amplitudes and orientations evolve as the oscillation propagates through the disk.

\begin{figure*}
\begin{center}
\includegraphics[width=1.0\textwidth]{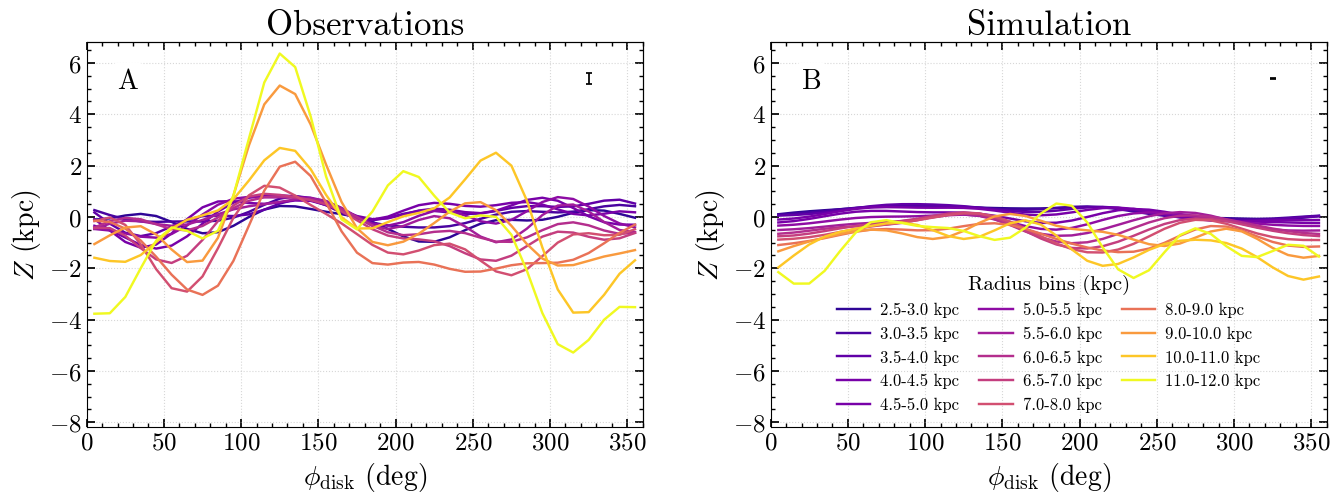}
\caption{Fourier-reconstructed vertical displacement of the LMC disk as a function of azimuth in the observations and simulation.
    (A) Observed and (B) simulated vertical displacement $Z$ reconstructed as a function of disk-plane azimuth $\phi_{\rm disk}$. Each colored curve represents a radial annulus, as indicated in the legend, from the inner disk (dark purple) to the outer periphery (yellow). The representative error bar in the upper right corner of each panel indicates the median propagated $1\sigma$ uncertainty in the reconstructed $Z$ field, calculated using the full covariance matrix of the Fourier fit.}
\label{fig:z_vs_pa}
\end{center}
\end{figure*}


At the nominal present-day epoch, approximately 140~Myr after the collision, the outer LMC disk is displaced toward the observer, producing predominantly positive $Z$ in the disk periphery, opposite in sign to the observed axisymmetric offset. As the simulation evolves, the vertical oscillation changes phase and the outer-disk displacement reverses sign approximately 120~Myr later.

To identify the snapshot most directly comparable to the observations, we computed the root-mean-square (RMS) difference between the observed and simulated Fourier-reconstructed $Z$ maps (Section~\ref{sec:fourier_comp}) at every snapshot of the simulation, over the 648 wedges between $R=2$ and 12~kpc used for the maps in Section~\ref{sec:fourier_comp}. The RMS is lowest ($\sim$1.35~kpc) at the snapshot approximately 260~Myr after the recent collision, compared with $\sim$2.0~kpc at the nominal present-day epoch. We therefore adopt this snapshot for all comparisons with the observations. Because snapshots are saved every 20~Myr, its time is determined only to within that cadence.


We do not interpret this time as a measurement of the time since the actual LMC--SMC collision. The sign of the warp at the nominal present day depends on the LMC model: in \citet{Garver2026}, the fiducial model shows a positive warp at that epoch, whereas the models with a thicker disk and with a more diffuse dark matter halo show a negative one, although those two models also reach $L_{\rm MS}=0^{\circ}$ earlier than the fiducial model, by 80 and 40~Myr, respectively. The 260~Myr should therefore be regarded as the time of best match for the fiducial model rather than a model-independent measurement of the collision time. Independent age constraints do not distinguish between the two times: young stars and clusters in the Magellanic Bridge have ages of up to $\sim$300~Myr \citep{Harris2009,JacyszynDobrzeniecka2016,Bitsakis2018}, consistent with an interaction either $\sim$140 or $\sim$260~Myr ago. In the following section, we apply the same Fourier decomposition to the observed and simulated vertical-displacement fields, allowing us to determine which components of the observed LMC structure are reproduced by the recent-collision model.

\section{Fourier Decomposition and Observation-Simulation Comparison}
\label{sec:fourier_comp}

In this section, we quantify and compare the vertical structure of the observed and simulated LMC disks. We decompose the vertical displacement field into low-order azimuthal Fourier modes within radial annuli, allowing us to trace the radial evolution of the perturbation and isolate its dominant components.

\subsection{Fourier method}
\label{sec:fourier_method}

We model the vertical displacement field as a Fourier series in disk-plane azimuth $\phi_{\rm disk}$ within fixed radial annuli:

\begin{equation}
\begin{aligned}
Z(R,\phi_{\rm disk}) = A_0(R)
&+ \sum_{m=1}^{N}
\left[
B_m(R)\cos(m\phi_{\rm disk}) \right. \\
&\left. + C_m(R)\sin(m\phi_{\rm disk})
\right],
\end{aligned}
\label{eq:fourier_series}
\end{equation}

\noindent where $A_0(R)$ represents the azimuthally averaged vertical displacement of the disk midplane at radius $R$, and the higher-order terms describe non-axisymmetric vertical structure. In this framework, the $m=1$ mode traces an integral-sign or S-shaped warp, while higher-order modes describe progressively more complex azimuthal structure. We retain modes up to $N=3$ throughout the analysis; tests including higher-order modes do not significantly change the dominant low-order structure.

The model is fit independently in radial annuli using bins of width $\Delta R = 0.5$~kpc for $2 \leq R < 7$~kpc and $\Delta R = 1$~kpc for $7 \leq R \leq 12$~kpc. Annuli containing fewer than 10 spatial bins are excluded from the fit. For each annulus, we use the median properties of the tangent-plane hybrid bins described in Section~\ref{sec:hybrid_binning} to define the representative disk-frame position and observed heliocentric distance.

To reduce biases associated with line-of-sight distance scatter, we fit the Fourier model in distance space rather than directly in $Z$. For a trial set of Fourier coefficients, the model vertical displacement field is evaluated at the representative disk-plane position of each bin. The model point $(X,Y,Z_{\rm model})$ is then transformed back into the sky-plane coordinate system using the adopted inclination and line-of-nodes position angle, and converted into a predicted heliocentric distance $D_{\rm model}$. The residuals minimized are

\begin{equation}
r_j =
\frac{D_{{\rm obs},j} - D_{{\rm model},j}}
{\sigma_{D,j}},
\label{eq:distance_residuals}
\end{equation}

\noindent where $D_{{\rm obs},j}$ is the observed median distance of spatial bin $j$ and $\sigma_{D,j}$ is a robust estimate of the uncertainty on the median distance of that bin. We estimate $\sigma_{D,j}$ from the median absolute deviation of the stellar distances within each bin, scaled as the uncertainty on the median. The residual vector is minimized using robust nonlinear least-squares optimization.

Formal $1\sigma$ uncertainties on the fitted Fourier coefficients are estimated from the full covariance matrix of the least-squares solution, including correlations between fitted parameters. For the non-axisymmetric modes, we define the Fourier amplitudes as

\begin{equation}
A_m(R) = \sqrt{B_m^2(R) + C_m^2(R)} \quad (m\geq1).
\label{eq:fourier_amplitude}
\end{equation}

The uncertainties on these amplitudes are computed by propagating the covariance between the corresponding sine and cosine coefficients. We also propagate the full covariance matrix into the reconstructed vertical displacement field. Details of the uncertainty estimation and propagation are given in Appendix~\ref{app:B}.

\begin{figure*}
\begin{center}
\includegraphics[width=1.0\textwidth]{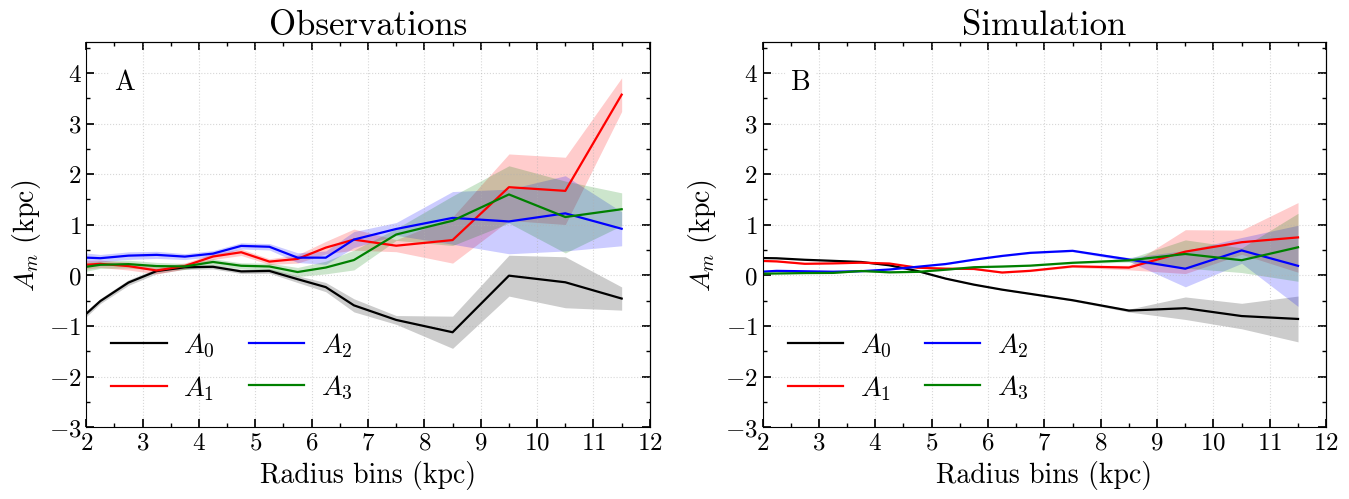}
\caption{Fourier mode amplitudes of the LMC vertical structure as a function of radius in the observations and simulation.
    (A) Observed and (B) simulated Fourier mode amplitudes $A_m$ ($m=0$--3) as a function of galactocentric radius. Shaded bands indicate propagated $1\sigma$ uncertainties calculated using the full covariance matrix of the least-squares fit. For $m\geq1$, these uncertainties include the covariance between the fitted sine and cosine coefficients. The $m=0$ term $A_0$ (black) represents the azimuthally averaged vertical offset of the disk midplane, while higher-order modes describe non-axisymmetric structure on progressively smaller azimuthal scales.}
\label{fig:fourier_moments}
\end{center}
\end{figure*}

The simulation is analyzed with the same definitions as the observations. The simulated LMC star particles are transformed into the disk frame using the same fiducial distance $D_0$, inclination $i$, and line-of-nodes position angle $\theta$ (Section~\ref{sec:coordinate_transformation}), with the same definition of $\phi_{\rm disk}$, and the Fourier model of Equation~\ref{eq:fourier_series} is fitted in the same radial annuli. No rotation or other alignment of the simulated disk is adjusted to match the observed warp, so the orientations $\phi_1$ of the observed and simulated $m=1$ terms can be compared directly.

\subsection{Azimuthal vertical profiles}
\label{sec:fourier_azimuthal_profiles}

Figure~\ref{fig:z_vs_pa} shows the Fourier-reconstructed vertical displacement $Z$ as a function of disk-plane azimuth $\phi_{\rm disk}$ for radial annuli spanning 2.0 to 12~kpc, comparing the observations (Panel A) and the simulation (Panel B). The representative error bar in each panel shows the median propagated uncertainty on the reconstructed $Z$ field, computed from the full covariance matrix of the Fourier fit.

In the observations, the inner annuli ($R \lesssim 5$~kpc; purple curves) exhibit low-amplitude azimuthal variation confined within $\pm$1~kpc of the disk midplane. At intermediate radii ($R\approx5$--9~kpc; magenta to salmon curves), $Z$ is predominantly negative, with typical values of $-1$ to $-2$~kpc and minima of approximately $-3$~kpc near $\phi_{\rm disk}\approx60$--$75^{\circ}$. The exception is a positive feature at $\phi_{\rm disk}\approx100$--$150^{\circ}$, which grows with radius to approximately $+2$~kpc in the 8--9~kpc annulus. Beyond 9~kpc (orange to yellow curves), this feature dominates and the profile becomes strongly asymmetric: $Z$ reaches approximately $+6$~kpc near $\phi_{\rm disk}\approx125^{\circ}$ in the 11--12~kpc annulus, with minima approaching $-5$~kpc near $\phi_{\rm disk}\approx315^{\circ}$.

The simulation shows the same radial progression. In the inner disk ($R \lesssim 5$~kpc), the profile shows little azimuthal variation and lies within $\pm$0.5~kpc of the disk midplane. At intermediate radii ($R\approx5$--9~kpc), $Z$ becomes progressively more negative with radius, reaching approximately $-1$~kpc. In the outer disk ($R\gtrsim9$~kpc), the azimuthal variation increases and the profile becomes mildly asymmetric, with minima of approximately $-2.5$~kpc. Unlike the observations, however, the simulated profiles show no large positive excursion: $Z$ remains below approximately $+1$~kpc at all radii and azimuths.

In summary, the observations and the simulation agree in the radial progression of the vertical structure and in the sign and approximate magnitude ($\sim$1~kpc) of the negative displacement at intermediate radii. They differ in the outer disk. In the 11--12~kpc annulus, the observed profile spans approximately $-5$ to $+6$~kpc, a peak-to-peak range of $\sim$11~kpc, while the simulated profile spans approximately $-2.5$ to $+0.7$~kpc, a range of $\sim$3~kpc. The azimuthal variation in the simulation is therefore smaller than observed by a factor of $\sim$3.5. In the Fourier description, these two aspects correspond to the axisymmetric ($m=0$) offset and the $m=1$ warp, respectively, which we quantify as a function of radius in the following section.

\begin{figure*}
\begin{center}
\includegraphics[width=1.0\textwidth]{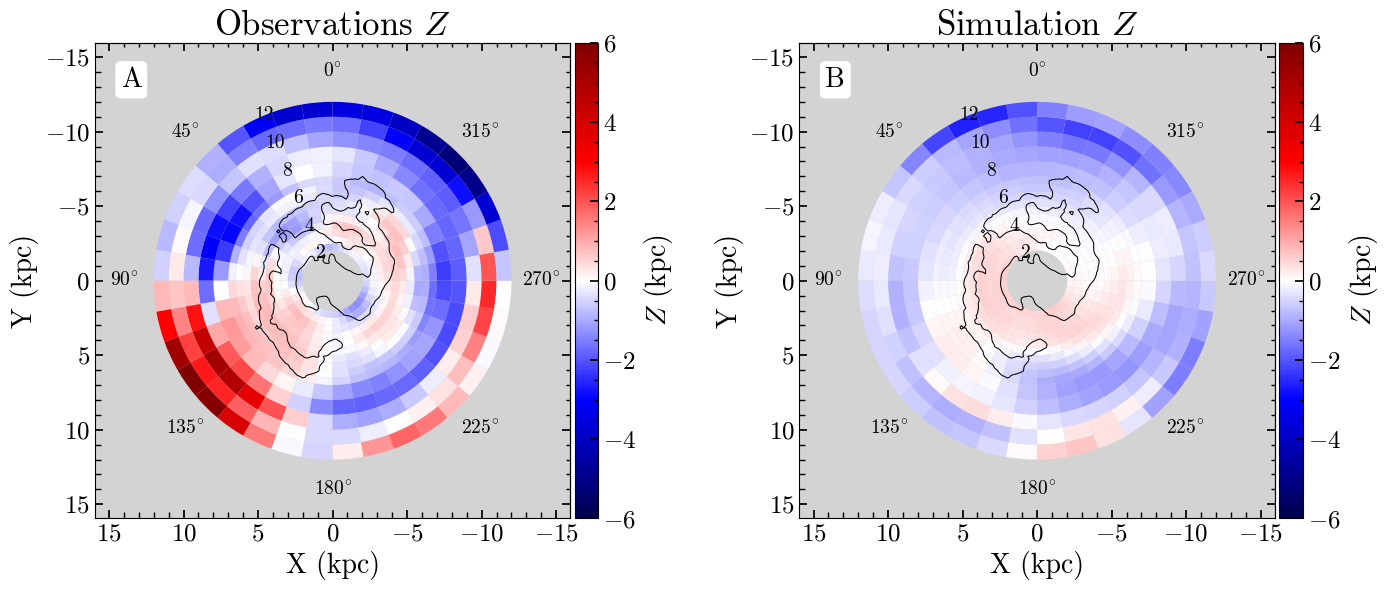}
\caption{Vertical displacement of the LMC disk in the observations and simulation.
    (A) Observed and (B) simulated Fourier-reconstructed vertical displacement $Z$ within each radial annulus and azimuthal wedge, shown in disk-plane coordinates. Blue and red indicate negative and positive displacements, respectively, relative to the best-fit disk plane. Black contours trace stellar overdensities associated with the LMC bar and spiral arm, identified using kernel density estimation \citep{JimenezArranz2023}. Disk-plane azimuths are labeled at $45^{\circ}$ intervals around the periphery, and galactocentric radii are marked at 2~kpc intervals.}
\label{fig:polarplot_Zp}
\end{center}
\end{figure*}

\subsection{Radial Fourier amplitudes}
\label{sec:fourier_moments}

Figure~\ref{fig:fourier_moments} shows the Fourier amplitudes up to $m = 3$ as a function of disk-plane galactocentric radius for the observations (Panel A) and the simulation (Panel B). The axisymmetric term $A_0$ is signed, while the amplitudes $A_m$ for $m\geq1$ are non-negative by definition (Equation~\ref{eq:fourier_amplitude}). We describe the profiles in three radial regimes.

In the inner disk ($R\lesssim5$~kpc), all terms are small in both the observations and the simulation. In the observations, the non-axisymmetric amplitudes remain below $\sim$0.6~kpc ($A_1\lesssim0.45$~kpc, $A_2\approx0.35$--0.6~kpc, and $A_3\approx0.2$~kpc), and $A_0$ lies within $0.2$~kpc of zero between 3 and 5~kpc. Inside $\sim$3~kpc, $A_0$ decreases to approximately $-0.75$~kpc in the innermost annulus, where the bar region contributes (Section~\ref{sec:extinction_correction}). In the simulation, all terms remain within $\sim$0.40~kpc of zero over the same radial range.

At intermediate radii ($R\approx5$--9~kpc), the axisymmetric term grows in both cases (black curves). In the observations, $A_0$ decreases steadily from approximately zero at 5~kpc to $-1.2\pm0.25$~kpc at $R\approx8.5$~kpc. Quoted uncertainties are the formal $1\sigma$ values described in Appendix~\ref{app:B}. In the simulation, $A_0$ also changes sign near 5~kpc and decreases to $-0.7$~kpc at $R\approx8.5$~kpc, about 60\% of the observed value. The two differ in the non-axisymmetric terms. In the observations, $A_2$ and $A_3$ grow to $\sim$1~kpc at $R\approx7.5$--8.5~kpc, comparable to $|A_0|$, while $A_1$ remains at 0.5--0.7~kpc. In the simulation, all non-axisymmetric amplitudes remain below $\sim$0.5~kpc.

In the outer disk ($R\gtrsim9$~kpc), the observed structure is dominated by the $m=1$ term. $A_1$ rises from $\sim$0.6~kpc at 8.5~kpc to $\sim$1.7~kpc at 9.5--10.5~kpc and $3.6\pm0.3$~kpc at $R\approx11.5$~kpc, where it exceeds every other term by a factor of more than 2.5. At 9.5--10.5~kpc, $A_1$ is the largest term, but $A_2$ and $A_3$ are comparable (1.1--1.6~kpc) within the uncertainties. Over the same range, the observed $A_0$ is consistent with zero at 9.5--10.5~kpc and is approximately $-0.45$~kpc at 11.5~kpc. In the simulation, $A_1$ also begins to rise beyond $\sim$9~kpc, but reaches only $\sim$0.45~kpc at 9.5~kpc and $\sim$0.75~kpc at 11.5~kpc, smaller than observed by factors of $\sim$4--5. The simulated $A_0$ remains negative, at approximately $-0.70$ to $-0.85$~kpc, and is the largest term at these radii, while the $m\geq2$ amplitudes remain below $\sim$0.55~kpc.

In summary, the observations and the simulation agree in three respects: both are nearly flat inside $\sim$5~kpc; in both, $A_0$ becomes negative beyond $\sim$5~kpc and reaches a comparable value at 8.5~kpc ($-1.2$ and $-0.7$~kpc, respectively); and in both, $A_1$ begins to grow beyond $\sim$9~kpc. They differ in three respects: the simulated $A_1$ is a factor of $\sim$4--5 smaller than observed in the outer disk; the observed $m=2$ and $m=3$ amplitudes of $\sim$1~kpc beyond $\sim$7~kpc have no counterpart in the simulation; and the observed $A_0$ returns toward zero beyond 9~kpc, while the simulated $A_0$ does not.

\subsection{Reconstructed field and isolated component maps}
\label{sec:fourier_z_map}


\begin{figure*}
\begin{center}
\includegraphics[width=1.0\textwidth]{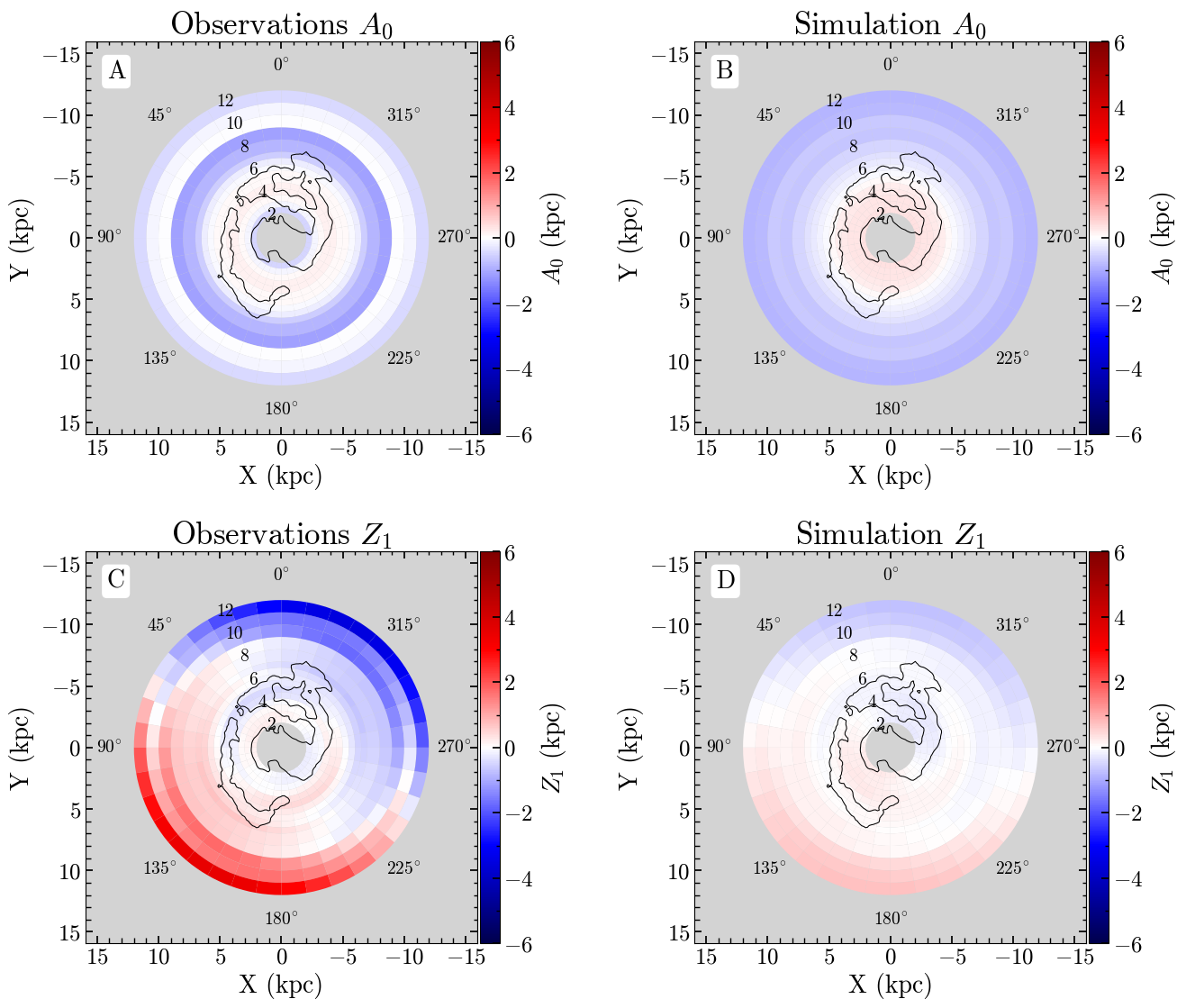}
\caption{Spatial maps of the two lowest-order Fourier terms of the LMC vertical displacement field (Equation~\ref{eq:fourier_series}) in the observations and the simulation.
    (A) Observed and (B) simulated axisymmetric term $A_0(R)$, the azimuthally averaged vertical offset, which is constant around each annulus.
    (C) Observed and (D) simulated $m=1$ term, $Z_1(R,\phi_{\rm disk}) = A_1(R)\cos\left[\phi_{\rm disk}-\phi_1(R)\right]$ (Equation~\ref{eq:z1}), where $A_1$ is the amplitude shown in Figure~\ref{fig:fourier_moments} and $\phi_1$ is the azimuth of maximum positive displacement. Around each annulus, $Z_1$ varies between $+A_1$ and $-A_1$, producing a dipole of positive and negative displacement on opposite sides of the disk.
    All panels are shown in disk-plane coordinates $(X,Y)$ on the same color scale, with positive values corresponding to displacement toward the observer. Black contours trace stellar overdensities associated with the LMC bar and spiral arm \citep{JimenezArranz2023}. Disk-plane azimuths are labeled at $45^{\circ}$ intervals around the periphery, and galactocentric radii are marked at 2~kpc intervals.}
\label{fig:4Panel_polarplot_A0_Zp1}
\end{center}
\end{figure*}

Figure~\ref{fig:polarplot_Zp} shows the full Fourier reconstruction of $Z$ across the LMC disk plane in the observations (Panel A) and the simulation (Panel B), with black contours tracing the bar and spiral arm of the LMC \citep[see][for details of the contours]{JimenezArranz2023}. Disk-plane azimuths are labeled at $45^{\circ}$ intervals around the periphery, and galactocentric radii are marked at 2~kpc intervals. Figure~\ref{fig:4Panel_polarplot_A0_Zp1} isolates the two lowest-order terms of Equation~\ref{eq:fourier_series}: the axisymmetric offset $A_0(R)$, which is constant around each annulus, and the $m=1$ term,

\begin{equation}
\begin{aligned}
Z_1(R,\phi_{\rm disk}) &= B_1(R)\cos\phi_{\rm disk} + C_1(R)\sin\phi_{\rm disk} \\
&= A_1(R)\cos\left[\phi_{\rm disk}-\phi_1(R)\right],
\end{aligned}
\label{eq:z1}
\end{equation}

\noindent where $A_1$ is the amplitude defined in Equation~\ref{eq:fourier_amplitude} and   $\phi_1=\operatorname{atan2}(C_1,B_1)$, the four-quadrant arctangent of $C_1/B_1$, is the azimuth of maximum positive displacement. $A_1$ therefore measures the size of the warp and $\phi_1$ its orientation.

\begin{figure*}
\begin{center}
\begin{interactive}{animation}{full_lmc_smc_vdb_movie_shorter_credits_apj.mp4}
\includegraphics[width=0.90\textwidth]{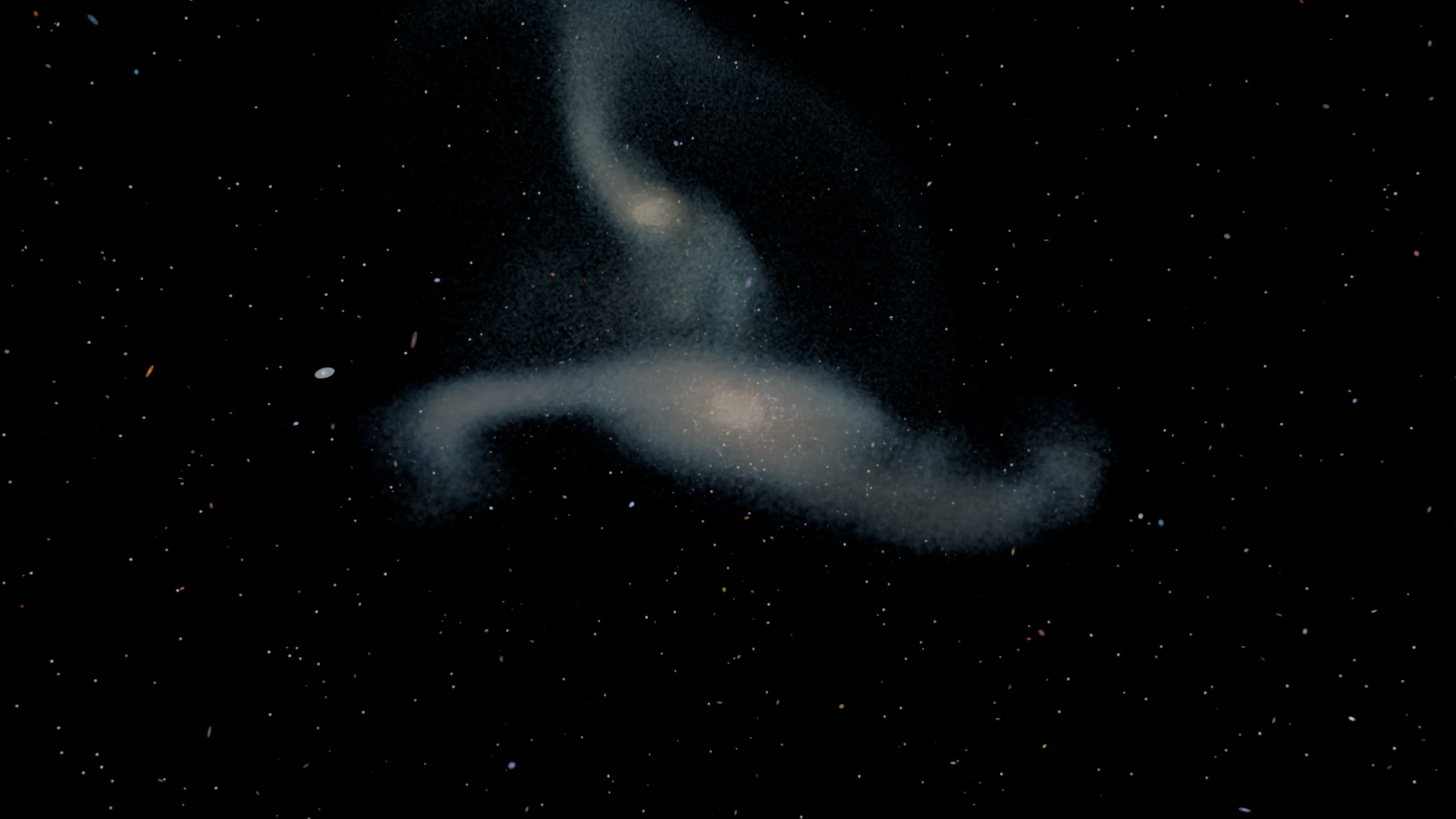}
\end{interactive}
\caption{Volume rendering of the $N$-body simulation of the LMC--SMC interaction \citep{Garver2026} at the snapshot that best matches the observed LMC, showing the warped LMC disk (lower center, nearly edge-on) and the SMC (above). Colors and background objects are illustrative. This figure is a frame from a 40~s animation, available in the online journal, which follows the simulation through the collision and then orbits the final configuration to show the warp from all viewing angles. The animation has no annotations.}
\label{fig:sim_render}
\end{center}
\end{figure*}

The observed map (Figure~\ref{fig:polarplot_Zp}A) shows the three radial regimes identified in Sections~\ref{sec:fourier_azimuthal_profiles} and \ref{sec:fourier_moments}. Inside $\sim$5~kpc, $|Z|\lesssim1$~kpc. At intermediate radii ($R\approx5$--9~kpc), $Z$ is negative over most azimuths, with values of $-1$ to $-3$~kpc that are most negative near $\phi_{\rm disk}\approx45$--$90^{\circ}$; the exception is a positive region at $\phi_{\rm disk}\approx100$--$150^{\circ}$. Beyond $\sim$9~kpc, the map is dominated by two opposing lobes: a positive lobe centered near $\phi_{\rm disk}\approx125^{\circ}$, reaching $\sim+6$~kpc at $R\approx11$--12~kpc, and a negative lobe centered near $\phi_{\rm disk}\approx315^{\circ}$, reaching $\sim-5$~kpc. A weaker positive feature of $\sim+2$~kpc is also present at $\phi_{\rm disk}\approx200$--$270^{\circ}$ and $R\approx10$--12~kpc.

The simulated map (Figure~\ref{fig:polarplot_Zp}B) is slightly positive inside $\sim$5~kpc ($Z\lesssim0.5$~kpc) and negative at nearly all azimuths beyond it, with values of $-0.5$ to $-1$~kpc at $R\approx6$--9~kpc. In the outer disk, the simulated $Z$ is most negative, at $\sim-2.5$~kpc, over $\phi_{\rm disk}\approx300$--$45^{\circ}$, the same azimuths as the observed negative lobe. Positive values are confined to small regions of $\lesssim+0.7$~kpc near $\phi_{\rm disk}\approx150$--$200^{\circ}$, on the same side of the disk as the observed positive lobe but far weaker. The RMS difference between the two maps is 1.35~kpc (Section~\ref{sec:sims_morphology}), and is dominated by the outer disk, where the amplitudes of the observed lobes are not matched.

The axisymmetric maps (Figures~\ref{fig:4Panel_polarplot_A0_Zp1}A and B) show where the observations and the simulation agree and where they do not within these two dominant modes. In the observations, the negative offset forms a ring confined to $R\approx6$--9~kpc, reaching $-1.2$~kpc at 8--9~kpc; beyond it, $A_0$ returns to approximately zero at 9--11~kpc and is $-0.45$~kpc in the outermost annulus. In the simulation, $A_0$ becomes negative at the same radius ($\sim$5~kpc) and reaches $-0.7$~kpc at 8--9~kpc, about 60\% of the observed value, but it continues to decrease outward, to approximately $-0.85$~kpc at 11--12~kpc. The simulation therefore reproduces the sign, the onset radius, and the approximate magnitude of the observed offset at intermediate radii, but not its return toward zero in the outer disk.

The $m=1$ maps (Figures~\ref{fig:4Panel_polarplot_A0_Zp1}C and D) show a dipole in both the observations and the simulation, with positive displacement on one side of the disk and negative displacement on the opposite side. In the outer disk ($R\approx9$--10.5~kpc), the observed $m=1$ term has an amplitude of $A_1\approx1.7$~kpc and is oriented at $\phi_1\approx147$--$150^{\circ}$, with an uncertainty of $\sim$20$^{\circ}$ per annulus propagated from the covariance matrix of the fit (Appendix~\ref{app:B}): the disk is displaced toward the observer on the side centered at $\phi_{\rm disk}\approx148^{\circ}$ and away from the observer on the opposite side. At $R\approx6.5$--9~kpc the orientation is $\phi_1\approx90$--$110^{\circ}$, so the warp twists by $\sim$40$^{\circ}$ across $R\approx9$~kpc. The simulated $m=1$ term is much weaker, with $A_1\approx0.45$~kpc at 9.5~kpc. Over $R\approx6.5$--10~kpc, the simulated orientation agrees with the observed orientation to within $\sim$10$^{\circ}$ and follows the same twist with radius. At 9.5~kpc, however, the simulated amplitude is only marginally larger than its uncertainty, so its orientation is weakly constrained there and at larger radii. The $m=1$ warp is therefore present in the simulation with an orientation consistent with that observed, but with a substantially smaller amplitude.

The simulation reproduces some but not all of the observed vertical structure. It reproduces the flat inner disk, the negative axisymmetric displacement at intermediate radii, and the orientation of the $m=1$ warp. It does not reproduce the amplitude of the $m=1$ warp, the return of $A_0$ toward zero beyond $\sim$9~kpc, or the $m=2$ and $m=3$ structure (Section~\ref{sec:fourier_moments}) that sharpens the observed positive lobe and produces the weaker positive feature at $\phi_{\rm disk}\approx200$--$270^{\circ}$.

To help visualize the three-dimensional geometry of the perturbed disk, Figure~\ref{fig:sim_render} shows a volume rendering of the simulation at the snapshot compared with the observations, and the accompanying animation in the online journal follows the full interaction. The animation shows the SMC colliding with the LMC disk and the subsequent development of the large-scale vertical distortion, and then orbits the final configuration so that the warp can be viewed from all azimuths. The rendering is intended as a qualitative illustration; the quantitative comparison between the simulation and the observations is that of Figures~\ref{fig:polarplot_Zp} and \ref{fig:4Panel_polarplot_A0_Zp1}.

Together, these comparisons show that the recent LMC--SMC collision in this model produces a vertical response with the observed sign, radial onset, and orientation, but underpredicts the amplitude of its asymmetric component. The origin of this mismatch cannot be determined from a single simulation. However, \citet{Garver2026} find that the direction and amplitude of the warp differ substantially among models that use the same orbit but different LMC disk thicknesses and dark matter halo concentrations, which suggests that the $m=1$ amplitude is sensitive to the internal structure of the LMC. The collision geometry and dynamical effects not captured by the present simulation may also contribute.



\section{Discussion}
\label{sec:discussion}

\subsection{Robustness of the inferred vertical structure}
\label{sec:discussion_robustness}

The large-scale vertical structure identified in the outer LMC disk is robust against both statistical uncertainties and potential biases introduced by the disk-frame coordinate transformation. As discussed in Section~\ref{sec:hybrid_binning} and Appendix~\ref{app:A}, line-of-sight distance scatter can produce artificial vertical signals when the transformation is applied on a per-star basis. Our controlled tests show that binning in distance-independent tangent-plane coordinates $(\xi,\eta)$ prior to transforming into the disk frame strongly suppresses these spurious gradients, recovering median $Z$ values consistent with zero for a synthetic disk with no intrinsic vertical structure. This gives us confidence that the observed warp is not an artifact of the coordinate transformation.

The inferred morphology is also stable against sparse sampling in the low-density LMC periphery. We verified that the large-scale warp structure remains consistent across different minimum stellar-density thresholds adopted for the adaptive outer-disk bins. The same qualitative structure is visible in the heliocentric distance map (Figure~\ref{fig:Hybrid_binning}C), the azimuthal $Z(\phi_{\rm disk})$ profiles (Figure~\ref{fig:z_vs_pa}A), the radial Fourier amplitudes (Figure~\ref{fig:fourier_moments}A), and the reconstructed map (Figure~\ref{fig:polarplot_Zp}A). As described in Section~\ref{sec:fourier_method} and Appendix~\ref{app:B}, the propagated uncertainties from the Fourier fits remain smaller than the observed large-scale vertical amplitudes across most of the disk, with their spatial distribution shown in Figure~\ref{fig:polarplot_SigmaZp}. In particular, the two features on which our interpretation rests are statistically significant: the axisymmetric offset at intermediate radii, $A_0\approx-1.2\pm0.25$~kpc at 8.5~kpc, exceeds its $1\sigma$ uncertainty by a factor of $\sim$5 (black curve in Figure~\ref{fig:fourier_moments}A), and the $m=1$ amplitude exceeds its uncertainty by a factor of $\sim$2.5 at 9.5~kpc (red curve in Figure~\ref{fig:fourier_moments}A). The orientation of the $m=1$ term is also stable, at $\phi_1\approx147$--$150^{\circ}$ across the 9--10.5~kpc annuli (Section~\ref{sec:fourier_z_map}).

 
The comparison with the simulation supports a collision-driven origin for part of this structure. In the fiducial model, the collision produces a vertical response with the observed sign and radial onset of the axisymmetric offset, and with an $m=1$ orientation consistent with that observed, but it underpredicts the $m=1$ amplitude and does not reproduce the observed $m=2$ and $m=3$ structure (Section~\ref{sec:fourier_z_map}). This comparison is made at the snapshot that minimizes the RMS difference from the observed map, so the comparison epoch is selected using the observed morphology and is not an independent prediction (Section~\ref{sec:sims_morphology}). The existence of a large-scale vertical response does not, however, depend on the details of the fiducial model. \citet{Garver2026} re-simulated the same orbit with five alternative LMC models that vary the disk thickness, disk velocity dispersion, and dark matter halo concentration, and found vertical oscillations of the outer disk in most of them; the exception is the model with a more concentrated halo, which develops only small-scale ripples. The radius at which the perturbation appears depends on the disk thickness, from $\sim$5~kpc for the thinner disk to $\sim$8~kpc for the thicker disk, comparable to the radii at which the observed disk departs from the plane ($\sim$5~kpc for the axisymmetric offset and $\sim$9~kpc for the $m=1$ warp). The direction and amplitude of the warp, however, differ substantially among the models. A collision-driven vertical perturbation of the outer LMC disk is therefore a common outcome of this orbit, but its amplitude, sign, and azimuthal structure are model dependent.

\subsection{Comparison with previous LMC--SMC collision models}
\label{sec:discussion_previous_models}

We do not fit the geometry of the LMC--SMC collision in this work. The orbit of the simulation was selected by \citet{Garver2026} to match the present-day positions and velocities of the Clouds (Section~\ref{sec:sims_suite}), and its geometry is described in Section~\ref{sec:sims_orbit}: the most recent encounter reaches a minimum separation of $\sim$3~kpc, $\sim$140~Myr before the nominal present day, with the SMC trajectory inclined by $\sim$83$^{\circ}$ to the normal of the LMC disk. The only quantity we adjust using the observed vertical structure is the time of the comparison: the RMS difference between the observed and simulated $Z$ maps is lowest $\sim$260~Myr after the collision (Section~\ref{sec:sims_morphology}). Here we compare these properties with previous models and observational constraints.

With a minimum separation of $\sim$3~kpc, the simulation is a direct-collision model, in the sense that the impact parameter is smaller than the radius of the LMC stellar disk \citep{Zivick2018}. The impact parameter and collision time of this simulation are close to those of previous work. In Model~2 of \citet{Besla2012}, the SMC collides with the LMC with an impact parameter of $\sim$2~kpc about 100~Myr ago \citep[see also][]{Besla2016}. From analytic orbit modeling based on the proper motions of the Clouds, \citet{Zivick2018} found a most probable collision time of $147\pm33$~Myr ago and an impact parameter of $7.5\pm2.5$~kpc. \citet{Choi2022} compared the residual proper-motion field of the LMC disk with simulated LMCs based on the \citet{Besla2012} models and concluded that a direct collision is required to reproduce the observed disk heating and kinematic asymmetry, with an impact parameter of $\leq$10~kpc and a collision time of $<$250~Myr; adopting a collision time of $\sim$140--160~Myr, they inferred an impact parameter of $\sim$5~kpc. Simulations of the LMC bar and of the structure of the SMC favor impact parameters of $\sim$2~kpc and collision times of $\sim$100--200~Myr \citep{Rathore2025,Rathore2026}. The impact parameter and collision time of the simulation used here ($\sim$3~kpc and $\sim$140~Myr before the nominal present day) lie within these ranges.


The time of best match to the observed vertical structure, $\sim$260~Myr after the collision, is later than all of these estimates, and slightly beyond the limit of $<$250~Myr that \citet{Choi2022} derived from the kinematics of the LMC disk. We do not regard it as a revised estimate of the collision time, because the orbit of the simulation matches the observed positions and velocities of the Clouds only at the nominal present day, $\sim$140~Myr after the collision. We instead interpret the difference as a phase offset: the observed outer disk is at a later phase of its vertical oscillation than the fiducial model predicts for the present day. That phase depends on the structure of the LMC as well as on the time since the collision (Section~\ref{sec:sims_morphology}). We nevertheless use the fiducial model throughout, since the orbit was not refitted for the alternative LMC models, and selecting an LMC model by the sign of its warp would use the quantity being tested.

Previous work has constrained the collision using the Magellanic Bridge, the distribution of stripped MC material, the LMC bar, or the in-plane kinematic response of the LMC disk \citep[e.g.,][]{Besla2012,Besla2016,Zivick2019,Cheng2022,Choi2022,Rathore2025,Rathore2026}. Our comparison uses a different observable, the vertical structure of the outer LMC disk. In the fiducial direct-collision model, the collision excites a vertical response with the observed sign and radial onset of the axisymmetric offset and with an $m=1$ orientation consistent with that observed (Section~\ref{sec:fourier_z_map}). We have not tested orbits with wider passages, so our comparison does not by itself exclude them. That case has been examined independently by \citet{Rathore2026b}, who analyzed a simulation of the LMC--SMC pair without the MW, in which the SMC completes four pericentric passages with separations decreasing from $\sim$35 to $\sim$7.5~kpc and no direct collision occurs. They found that the outer LMC disk ($R\approx10$~kpc) reaches a mean vertical deviation of $\lesssim$1~kpc, compared with the observed vertical deviation of at least 2~kpc at $R\approx7$--10~kpc that they adopt, and concluded that an orbit with a closest approach of $\geq$7~kpc does not produce large enough vertical perturbations, so that a much closer encounter, such as a direct collision, is needed.

The same study bears on the amplitude mismatch found here. \citet{Rathore2026b} showed that the LMC disk should already have been warped, with mean vertical deviations of up to $\sim$1~kpc, before the recent collision, as a result of earlier LMC--SMC interactions and of the torques exerted by the distorted LMC dark matter halo. The simulation used here begins 2.5~Gyr ago with the LMC disk in equilibrium and includes only one earlier passage, so it does not contain distortions accumulated over a longer interaction history. Such pre-existing warps may contribute to the amplitude of the observed $m=1$ warp, which our direct-collision model underpredicts (Section~\ref{sec:fourier_z_map}). Testing this requires simulations that follow both the long-term interaction history and the recent collision.

\subsection{The observed--simulated $m=1$ amplitude discrepancy}
\label{sec:discussion_obs_sim_discrepancy}

Although the simulation reproduces several features of the observed vertical structure of the LMC, it underpredicts the amplitude of the outer $m=1$ warp. In the observations, $A_1$ rises beyond $R\sim9$~kpc to $\sim$1.7~kpc at 9.5--10.5~kpc and $\sim$3.6~kpc at $R\approx11.5$~kpc (red curve in Figure~\ref{fig:fourier_moments}A). In the simulation, $A_1$ also rises beyond $\sim$9~kpc, but reaches only $\sim$0.45~kpc at 9.5~kpc and $\sim$0.75~kpc at 11.5~kpc (red curve in Figure~\ref{fig:fourier_moments}B), smaller than observed by factors of $\sim$4--5. At the same time, the simulation reproduces the sign and radial onset of the axisymmetric offset: $A_0$ becomes negative beyond $\sim$5~kpc in both cases and reaches approximately $-1.2$~kpc in the observations and approximately $-0.7$~kpc in the simulation at 8.5~kpc (black curves in Figures~\ref{fig:fourier_moments}A and B). It also reproduces the orientation of the $m=1$ warp, which agrees with that observed to within $\sim$10$^{\circ}$ over $R\approx6.5$--10~kpc (Figures~\ref{fig:4Panel_polarplot_A0_Zp1}C and D; Section~\ref{sec:fourier_z_map}). The principal discrepancy is therefore in the amplitude of the asymmetric component, not in its presence or orientation.

The origin of this discrepancy cannot be determined from a single simulation, but several factors may contribute. The first is the internal structure of the LMC: \citet{Garver2026} find that the direction and amplitude of the warp differ substantially among models that use the same orbit but different disk thicknesses and dark matter halo concentrations (Section~\ref{sec:sims_morphology}). The second is the collision geometry and the mass of the SMC, which we have not varied. In the simulation, the SMC trajectory is inclined by $\sim$83$^{\circ}$ to the normal of the LMC disk and is therefore nearly in the plane of the disk; for such a geometry, the amplitude of the vertical response may be sensitive to modest changes in the trajectory. The third is vertical structure accumulated over a longer interaction history than the simulation follows: the simulation begins 2.5~Gyr ago with the LMC disk in equilibrium and includes only one earlier passage, whereas \citet{Rathore2026b} find that earlier LMC--SMC interactions and the resulting halo torques warp the disk by up to $\sim$1~kpc before the recent collision (Section~\ref{sec:discussion_previous_models}).

The amplitude mismatch therefore identifies the aspect of the model that the observations constrain most strongly. The agreement in the sign and radial onset of the axisymmetric offset and in the orientation of the $m=1$ warp shows that a recent direct collision can produce vertical structure of the observed form. The amplitude of the asymmetric component, together with the return of $A_0$ toward zero beyond $\sim$9~kpc and the observed $m=2$ and $m=3$ structure (Section~\ref{sec:fourier_moments}), is not reproduced and provides a target for future models.

A quantitative constraint on the structure of the LMC would require applying the same decomposition to simulations that span disk thickness and halo concentration, with the orbit refitted for each model, which we defer to future work.

\subsection{Implications for dwarf--dwarf interactions}
\label{sec:discussion_implications}

Our results support the interpretation that the large-scale vertical structure of the LMC disk is a dynamical response to the recent interaction between the MCs. Together, the observations and the simulation provide a rare opportunity to study the aftermath of a dwarf--dwarf collision in a nearby resolved galaxy. While galaxy interactions are widely recognized as a fundamental driver of galaxy evolution, their detailed dynamical impact on stellar disks is typically inferred from semi-resolved and unresolved systems or from numerical simulations. The MCs instead provide a uniquely accessible laboratory in which the global three-dimensional response of a dwarf disk can be mapped directly.

In the LMC, the vertical structure consists of an azimuthally averaged displacement of the disk at intermediate radii and a large-scale S-shaped warp across the stellar periphery. In the simulation, a direct collision with the SMC excites coherent, galaxy-scale bending of the LMC disk that persists for several hundred Myr after the interaction. The simulation reproduces the sign and radial onset of the observed offset and the orientation of the observed warp, although not its amplitude, which supports the interpretation that these structures were generated by the recent LMC--SMC collision.

More broadly, these results suggest that vertical perturbations induced by satellite interactions may be a common feature of dwarf galaxy evolution. Because dwarf galaxies are the most numerous galaxies in the Universe and are thought to represent the fundamental building blocks of hierarchical galaxy assembly, repeated dwarf--dwarf encounters are expected to occur frequently throughout cosmic time. The LMC therefore provides a nearby observational example of how such interactions can restructure stellar disks on galactic scales, offering an important framework for interpreting warped outer disks and disturbed stellar peripheries in galaxies throughout the nearby Universe.

\section{Summary}
\label{sec:conclusion}

We have used the \textit{Gaia} DR3 red clump distance map of the LMC from \citetalias{Oden2026} to reconstruct the three-dimensional structure of its stellar disk and to compare the observed vertical displacement field with an $N$-body simulation of the recent LMC--SMC collision from \citet{Garver2026}. To minimize biases introduced by line-of-sight distance scatter, we measured median distances in distance-independent tangent-plane coordinates $(\xi,\eta)$ before transforming into the LMC disk frame. We then quantified the resulting vertical structure using a low-order azimuthal Fourier decomposition in radial annuli, allowing the axisymmetric and non-axisymmetric components of the warp to be measured separately.

The observed LMC disk is nearly flat in the inner disk ($R\lesssim5$~kpc) and shows coherent large-scale vertical structure beyond it. The dominant components are a negative axisymmetric ($m=0$) displacement at intermediate radii ($R\approx5$--9~kpc) and an $m=1$ integral-sign warp in the outer disk ($R\gtrsim9$~kpc). The axisymmetric offset becomes negative beyond $\sim$5~kpc and reaches approximately $-1.2$~kpc near $R\approx8.5$~kpc. In the outer disk, the $m=1$ component grows rapidly, from $\sim$1.7~kpc at $R\approx9.5$--10.5~kpc to $\sim$3.6~kpc near $R\approx11.5$~kpc. These structures are significant relative to their statistical uncertainties and are not produced by sparse sampling in the LMC periphery or by the disk-frame coordinate transformation (Section~\ref{sec:discussion_robustness}).

We compared these measurements with a single simulation, the fiducial high-resolution model of \citet{Garver2026}, in which the SMC collides with the LMC with a minimum separation of $\sim$3~kpc, $\sim$140~Myr before the nominal present day, on a trajectory inclined by $\sim$83$^{\circ}$ to the normal of the LMC disk. In this simulation, the collision excites a vertical oscillation of the outer LMC disk with a period of $\sim$600~Myr. The RMS difference between the observed and simulated vertical displacement maps is lowest $\sim$260~Myr after the collision. This time depends on the adopted LMC model, and we do not regard it as a revised estimate of the collision time. At this snapshot, the simulation reproduces the sign and radial onset of the axisymmetric offset, with about 60\% of its observed magnitude at 8.5~kpc, and the orientation of the $m=1$ warp to within $\sim$10$^{\circ}$ over $R\approx6.5$--10~kpc. However, the simulated $m=1$ amplitude in the outer disk is smaller than observed by a factor of $\sim$4--5. The origin of this discrepancy cannot be determined from a single simulation. \citet{Garver2026} find that the direction and amplitude of the warp differ substantially among models that use the same orbit but different LMC disk thicknesses and dark matter halo concentrations, which suggests that the amplitude is sensitive to the internal structure of the LMC; the collision geometry and vertical structure that predates the collision may also contribute (Section~\ref{sec:discussion_obs_sim_discrepancy}). Together, these results support a collision-driven origin for the large-scale vertical structure of the LMC disk, and are consistent with the LMC being a nearby, resolved example of a dwarf disk still ringing from a recent dwarf--dwarf collision.

\section*{Acknowledgments}

S.J.O. and D.L.N. acknowledge support from National Science Foundation grants AST-1908331 and AST-2408159. S.J.O. also acknowledges support from the Montana Space Grant Consortium Graduate Fellowship under grant number 80NSSC20M0042. 

This work has made use of data from the European Space Agency (ESA) mission
{\it Gaia} (\url{https://www.cosmos.esa.int/gaia}), processed by the {\it Gaia}
Data Processing and Analysis Consortium (DPAC,
\url{https://www.cosmos.esa.int/web/gaia/dpac/consortium}). Funding for the DPAC has been provided by national institutions, in particular the institutions
participating in the {\it Gaia} Multilateral Agreement.

This work has also made use of data obtained with the Dark Energy Camera (DECam) at the Blanco 4m telescope at Cerro Tololo Inter-American Observatory. DECam was constructed by the Dark Energy Survey (DES) collaborating institutions: Argonne National Lab, University of California Santa Cruz, University of Cambridge, Centro de Investigaciones Energeticas, Medioambientales y Tecnologicas-Madrid, University of Chicago, University College London, DES-Brazil consortium, University of Edinburgh, ETH-Zurich, University of Illinois at Urbana-Champaign, Institut de Ciencies de l'Espai, Institut de Fisica d'Altes Energies, Lawrence Berkeley National Lab, Ludwig-Maximilians Universit\"at, University of Michigan, National Optical Astronomy Observatory, University of Nottingham, Ohio State University, University of Pennsylvania, University of Portsmouth, SLAC National Lab, Stanford University, University of Sussex, and Texas A\&M University. Funding for DES, including DECam, has been provided by the U.S. Department of Energy, National Science Foundation, Ministry of Education and Science (Spain), Science and Technology Facilities Council (UK), Higher Education Funding Council (England), National Center for Supercomputing Applications, Kavli Institute for Cosmological Physics, Financiadora de Estudos e Projetos, Funda\c{c}\~ao Carlos Chagas Filho de Amparo a Pesquisa, Conselho Nacional de Desenvolvimento Cientfico e Tecnol\'ogico and the Minist\'erio da Ci\^encia e Tecnologia (Brazil), the German Research Foundation-sponsored cluster of excellence "Origin and Structure of the Universe" and the DES collaborating institutions. The Cerro Tololo Inter-American Observatory, National Optical Astronomy Observatory is operated by the Association of Universities for Research in Astronomy (AURA) under a cooperative agreement with the National Science Foundation. 

The DELVE Collaboration gratefully acknowledges support from
Fermilab LDRD L2019-011, the NASA Fermi Guest Investigator Program Cycle 9 No.
91201, and the National Science Foundation under Grant No. AST-2108168,
AST-2108169, AST-2307126, and AST-2407526. This research is partially funded by a
generous gift from Charles Simonyi to the NSF Division of Astronomical Sciences. The
award is made in recognition of significant contributions to Rubin Observatory’s Legacy
Survey of Space and Time

Computational efforts were performed on the Tempest High Performance Computing System, operated and supported by University Information Technology Research Cyberinfrastructure (RRID:SCR\_026229) at Montana State University.

Portions of the text were refined with the assistance of OpenAI’s ChatGPT (GPT-5), which was used to improve clarity, grammar, and style.




\clearpage

\begin{figure*}
\begin{center}
\includegraphics[width=1.0\textwidth]{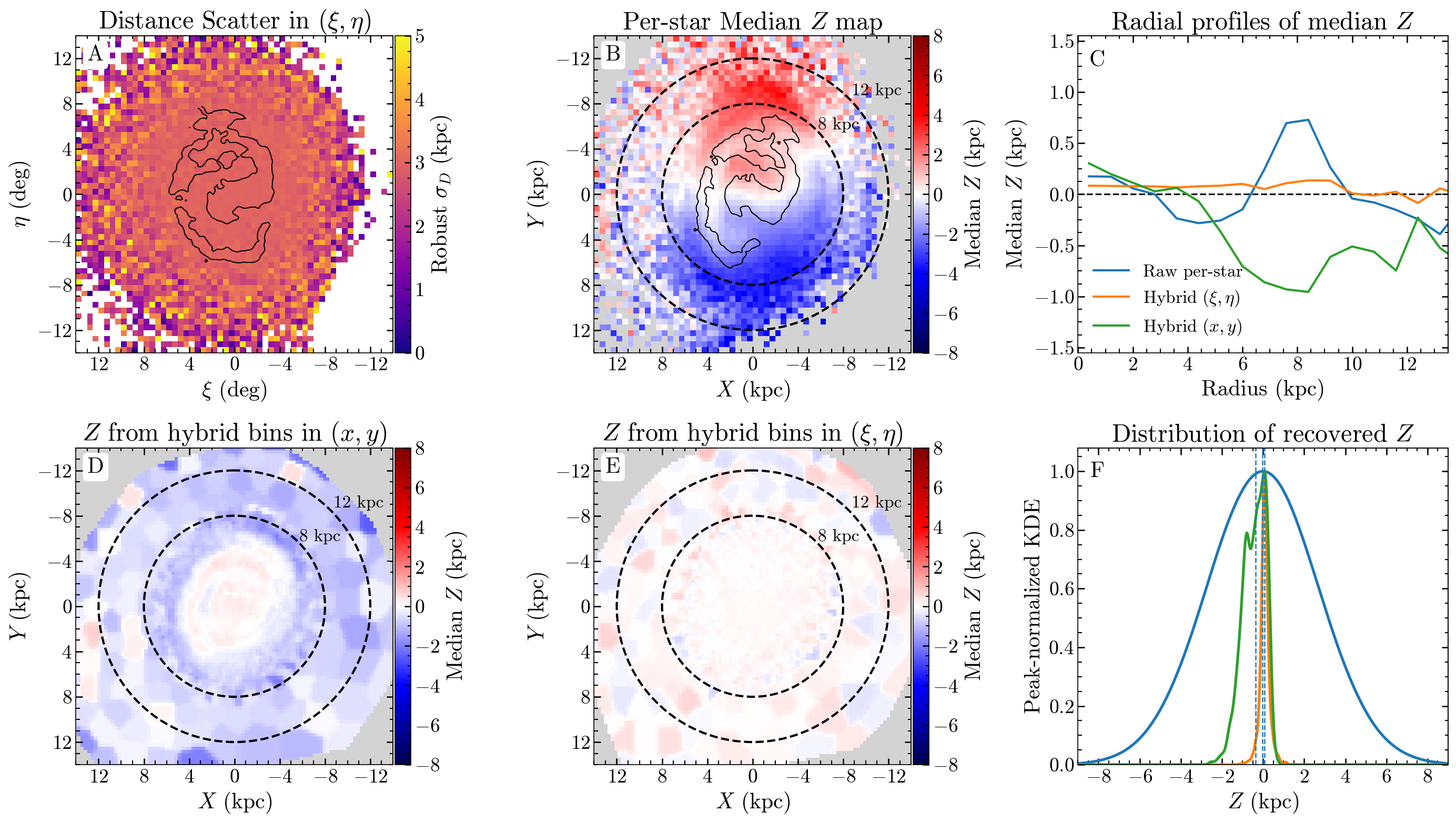}
\caption{Effect of line-of-sight distance scatter on the recovered disk-frame vertical structure.
    (A) Sky map of the robust dispersion of the mock distances after injecting Gaussian line-of-sight distance offsets.
    (B) Median $Z$ map obtained by transforming individual stars into the disk frame. The synthetic stars initially lie in an infinitesimally thin disk ($Z=0$) and are then assigned Gaussian line-of-sight distance offsets with a standard deviation of 3~kpc. The recovered map exhibits a coherent gradient reaching approximately $\pm4$~kpc, demonstrating that this procedure can produce spurious vertical structure in the presence of distance scatter.
    (C) Radial profiles of median $Z$ for three methods: per-star transformation (blue), hybrid binning in $(x,y)$ (green), and hybrid binning in $(\xi,\eta)$ (orange).
    (D) Median $Z$ map obtained using hybrid binning in $(x,y)$, which retains a systematic offset of approximately $-1$~kpc at large radii because distance scatter propagates into the binning coordinates.
    (E) Median $Z$ map obtained using hybrid binning in $(\xi,\eta)$, which suppresses the spurious structure and recovers values consistent with $Z=0$ across the disk.
    (F) Distributions of recovered $Z$, each normalized to its peak, showing that only binning in $(\xi,\eta)$ yields a distribution sharply centered on $Z=0$.}
\label{fig:Z_bias_test}
\end{center}
\end{figure*}

\twocolumngrid
\appendix

\section{Line-of-sight Distance Scatter and Artificial Vertical Structure}
\label{app:A}

A key systematic in measuring the vertical structure of the LMC disk arises from the combination of line-of-sight distance scatter and the disk-frame coordinate transformation. Because the LMC disk is inclined, an error in heliocentric distance does not project only into the line-of-sight direction; after rotation into the disk frame, it also produces an apparent displacement in the disk-frame vertical coordinate $Z$. If the transformation is applied to individual stars before spatial aggregation, these distance errors can therefore generate large spurious values of $|Z|$ that correlate with on-sky position. Such a signal can mimic a coherent warp even when no intrinsic vertical structure is present.

We test this effect using a controlled experiment based on the observed on-sky positions of the RC sample. Each star is first placed in an infinitesimally thin LMC disk with intrinsic $Z=0$, using the adopted LMC center, inclination and line-of-nodes position angle. This defines the heliocentric distance each star would have if it lay exactly in the reference disk plane. We then add Gaussian-distributed line-of-sight distance offsets with a standard deviation of 3~kpc to these synthetic distances, producing a mock catalog with realistic sky coverage and distance scatter but no intrinsic vertical structure. Because the input disk is exactly flat, any coherent spatial pattern recovered in $Z$ after applying the disk-frame transformation must be artificial.

Figure~\ref{fig:Z_bias_test}A shows the robust dispersion of the mock distances across the LMC footprint. Applying the disk-frame transformation directly to each star and then computing the median $Z$ field produces a strong coherent gradient with amplitudes reaching approximately $\pm4$~kpc, as shown in Figure~\ref{fig:Z_bias_test}B. This demonstrates that when individual distances set both the disk-frame positions and the bins, line-of-sight distance scatter produces artificial warp-like structure. The recovered signal is not caused by any physical vertical displacement, since the synthetic disk was constructed with $Z=0$ by design.

We next compare this per-star approach with two spatial-binning schemes. In the first, stars are binned in distance-dependent Cartesian coordinates $(x,y)$ before measuring the median vertical structure. This reduces the amplitude of the artificial signal relative to the per-star transformation, but does not fully remove the bias. As shown in Figure~\ref{fig:Z_bias_test}D, the recovered map retains a systematic negative offset of approximately $-1$~kpc at large radii. This residual bias occurs because the coordinates $(x,y)$ themselves depend on the noisy heliocentric distances; distance scatter therefore affects both the inferred vertical coordinate and the spatial coordinates used to define the bins.

In the second approach, stars are binned in tangent-plane coordinates $(\xi,\eta)$ before the transformation into the disk frame. These coordinates depend only on angular position on the sky and are independent of the measured heliocentric distance. For each bin, we compute the median heliocentric distance and then transform the bin into the disk frame using the median sky position and median distance. This procedure prevents individual distance errors from moving stars between spatial bins and avoids using noisy distance measurements to define the binning coordinates. As shown in Figure~\ref{fig:Z_bias_test}E, the recovered median $Z$ map remains centered near $Z=0$ across the disk, with no coherent warp-like gradient.

The radial profiles in Figure~\ref{fig:Z_bias_test}C and the recovered $Z$ distributions in Figure~\ref{fig:Z_bias_test}F summarize the comparison between the three approaches. The per-star transformation produces the largest artificial vertical signal, hybrid binning in $(x,y)$ reduces but does not eliminate the bias, and hybrid binning in $(\xi,\eta)$ recovers a distribution sharply centered on $Z=0$. This test therefore justifies our use of tangent-plane hybrid binning for the main analysis. Since the ($\xi$,$\eta$)-binned method recovers no coherent vertical structure when applied to a synthetic disk with known intrinsic $Z=0$, the large-scale warp measured in the real RC data is not produced by the coordinate transformation or by line-of-sight distance scatter alone.

\begin{figure*}
\begin{center}
	\includegraphics[width=0.75\textwidth]{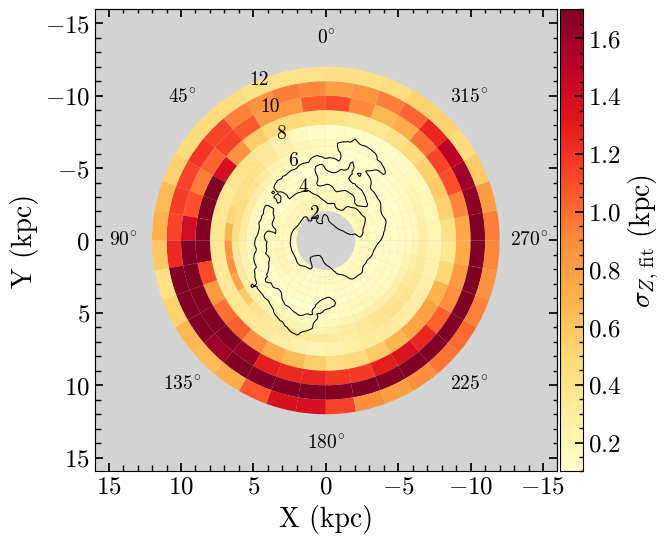} 
\caption{Propagated $1\sigma$ uncertainties in the Fourier-reconstructed vertical displacement field of the LMC.
    The map shows the per-bin uncertainty $\sigma_{Z,{\rm fit}}$ in disk-plane coordinates, calculated using the full covariance matrix of the Fourier least-squares fit. Median uncertainties are $\sim$0.17~kpc in the inner disk ($R<5$~kpc) and $\sim$0.51~kpc at larger radii ($R>5$~kpc), reaching $\sim$1.21~kpc in the annuli spanning $R\sim9$--11~kpc. Black contours trace stellar overdensities associated with the LMC bar and spiral structure \citep{JimenezArranz2023}. Disk-plane azimuths are labeled at $45^{\circ}$ intervals, and galactocentric radii are marked at 2~kpc intervals.}
	\label{fig:polarplot_SigmaZp}
\end{center}
\end{figure*}

\section{Fourier Uncertainty Estimation and Propagation}
\label{app:B}

Uncertainties on the fitted Fourier parameters are derived from the covariance matrix of the least-squares solution. For each radial annulus, the Fourier model is fitted to the distance-space residuals defined in Equation~\ref{eq:distance_residuals}. Let $\mathbf{J}$ be the Jacobian matrix of the residuals with respect to the fitted parameters, evaluated at the best-fitting solution. We estimate the parameter covariance matrix as

\begin{equation}
\mathbf{C} = s^2 \left(\mathbf{J}^{T}\mathbf{J}\right)^{-1},
\label{eq:covariance_matrix}
\end{equation}

\noindent where $s^2=\chi^2/{\rm dof}$ is the reduced variance of the fit. Here

\begin{equation}
\chi^2 = \sum_{j=1}^{N_{\rm bin}} r_j^2
= \sum_{j=1}^{N_{\rm bin}} \left(\frac{D_{{\rm obs},j} - D_{{\rm model},j}}{\sigma_{D,j}}\right)^2
\label{eq:chi2}
\end{equation}

\noindent is the sum of the squared residuals of Equation~\ref{eq:distance_residuals} at the best-fitting solution, and ${\rm dof} = N_{\rm bin} - p$ is the number of degrees of freedom, with $N_{\rm bin}$ the number of spatial bins in the annulus and $p=2N+1=7$ the number of fitted parameters. In cases where $\mathbf{J}^{T}\mathbf{J}$ is singular or poorly conditioned, we use a pseudo-inverse. The formal $1\sigma$ uncertainties on the fitted coefficients, including $A_0$ and the sine and cosine coefficients $(B_m,C_m)$, are given by the square roots of the diagonal elements of $\mathbf{C}$.

For the non-axisymmetric modes, the Fourier amplitude is defined as

\begin{equation}
A_m = \sqrt{B_m^2 + C_m^2} \quad (m \geq 1).
\end{equation}

We compute the uncertainty on $A_m$ by propagating the covariance between $B_m$ and $C_m$:

\begin{equation}
\sigma^2_{A_m} =
\nabla A_m^{T}\mathbf{C}_m\nabla A_m,
\label{eq:Fourier_amplitude_uncertainties}
\end{equation}

\noindent where $\mathbf{C}_m$ is the $2\times2$ covariance submatrix corresponding to $(B_m,C_m)$ and

\begin{equation}
\nabla A_m =
\left(
\frac{B_m}{A_m},
\frac{C_m}{A_m}
\right).
\end{equation}

This formulation accounts for correlations between the sine and cosine coefficients. In the limit $A_m\rightarrow0$, where the gradient becomes numerically unstable, we adopt

\begin{equation}
\sigma^2_{A_m} = {\rm Tr}(\mathbf{C}_m)
\end{equation}

\noindent as a stable approximation.

The orientation of the $m=1$ term, $\phi_1=\operatorname{atan2}(C_1,B_1)$ (Equation~\ref{eq:z1}), has an uncertainty, in radians, that follows from the same covariance submatrix:

\begin{equation}
\sigma^2_{\phi_1} = \frac{C_1^2\,{\rm Var}(B_1) + B_1^2\,{\rm Var}(C_1) - 2B_1C_1\,{\rm Cov}(B_1,C_1)}{A_1^4}.
\label{eq:phi1_uncertainty}
\end{equation}

\noindent For uncorrelated coefficients of equal variance, this reduces to $\sigma_{\phi_1}\approx\sigma_{A_1}/A_1$, so the orientation becomes poorly constrained as $A_1$ approaches its uncertainty.

We also propagate the full covariance matrix into the reconstructed vertical displacement field. For a given radial annulus, the uncertainty on the reconstructed model at azimuth $\phi$ is

\begin{equation}
\sigma_{Z,{\rm fit}}^2(\phi) =
\mathbf{G}(\phi)\mathbf{C}\mathbf{G}^{T}(\phi),
\label{eq:Z_uncertainty}
\end{equation}

\noindent where $\mathbf{G}(\phi)$ is the design vector of partial derivatives of $Z(\phi)$ with respect to the fitted Fourier parameters:

\begin{equation}
\mathbf{G}(\phi) =
\left[
1,
\cos\phi,
\sin\phi,
\cos(2\phi),
\sin(2\phi),
\ldots
\right].
\end{equation}

This provides a pointwise estimate of the uncertainty on the reconstructed vertical structure in each radial annulus. The resulting spatial distribution of $\sigma_{Z,{\rm fit}}$ across the LMC disk is shown in Figure~\ref{fig:polarplot_SigmaZp}.

The propagated uncertainties are small compared to the observed large-scale vertical structure across most of the LMC disk. As shown in Figure~\ref{fig:polarplot_SigmaZp}, the median reconstructed uncertainty is $\sim$0.17~kpc for $R<5$~kpc and $\sim$0.51~kpc for $R>5$~kpc. The uncertainties increase in the low-density outer disk and show a pronounced peak over $R\sim9$--11~kpc, where the median reaches $\sim$1.21~kpc. Although the uncertainties approach the signal amplitude in some outer annuli, the dominant $m=0$ and $m=1$ structures remain well constrained over the radial range in which the large-scale warp is measured.





\bibliography{ref} 

@ARTICLE{Toomre1972,
       author = {{Toomre}, Alar and {Toomre}, Juri},
        title = "{Galactic Bridges and Tails}",
      journal = {\apj},
         year = 1972,
        month = dec,
       volume = {178},
        pages = {623-666},
          doi = {10.1086/151823},
       adsurl = {https://ui.adsabs.harvard.edu/abs/1972ApJ...178..623T}
}

@ARTICLE{Barnes1992,
       author = {{Barnes}, Joshua E. and {Hernquist}, Lars},
        title = "{Dynamics of interacting galaxies.}",
      journal = {\araa},
         year = 1992,
        month = jan,
       volume = {30},
        pages = {705-742},
          doi = {10.1146/annurev.aa.30.090192.003421},
       adsurl = {https://ui.adsabs.harvard.edu/abs/1992ARA&A..30..705B}
}

@ARTICLE{Schweizer2000,
  author  = {{Schweizer}, Fran{\c c}ois},
  title   = {Interactions as a driver of galaxy evolution},
  journal = {Philosophical Transactions of the Royal Society of London Series A},
  year    = {2000},
  volume  = {358},
  pages   = {2063--2071},
  doi     = {10.1098/rsta.2000.0630},
  eprint  = {astro-ph/0002263},
  archivePrefix = {arXiv},
  primaryClass  = {astro-ph},
  adsurl = {https://ui.adsabs.harvard.edu/abs/2000RSPTA.358.2063S}
}

@ARTICLE{Huvsko2022,
       author = {{Hu{\v{s}}ko}, Filip and {Lacey}, Cedric G. and {Baugh}, Carlton M.},
        title = "{Statistics of galaxy mergers: bridging the gap between theory and observation}",
      journal = {\mnras},
         year = 2022,
        month = feb,
       volume = {509},
       number = {4},
        pages = {5918-5937},
          doi = {10.1093/mnras/stab3324},
archivePrefix = {arXiv},
       eprint = {2107.05601},
 primaryClass = {astro-ph.GA},
       adsurl = {https://ui.adsabs.harvard.edu/abs/2022MNRAS.509.5918H}
}

@ARTICLE{Amorisco2014,
       author = {{Amorisco}, N.~C. and {Evans}, N.~W. and {van de Ven}, G.},
        title = "{The remnant of a merger between two dwarf galaxies in Andromeda II}",
      journal = {\nat},
         year = 2014,
        month = mar,
       volume = {507},
       number = {7492},
        pages = {335-337},
          doi = {10.1038/nature12995},
archivePrefix = {arXiv},
       eprint = {1402.5142},
 primaryClass = {astro-ph.GA},
       adsurl = {https://ui.adsabs.harvard.edu/abs/2014Natur.507..335A}
}

@ARTICLE{Gomez2013,
       author = {{G{\'o}mez}, Facundo A. and {Minchev}, Ivan and {O'Shea}, Brian W. and {Beers}, Timothy C. and {Bullock}, James S. and {Purcell}, Chris W.},
        title = "{Vertical density waves in the Milky Way disc induced by the Sagittarius dwarf galaxy}",
      journal = {\mnras},
         year = 2013,
        month = feb,
       volume = {429},
       number = {1},
        pages = {159-164},
          doi = {10.1093/mnras/sts327},
archivePrefix = {arXiv},
       eprint = {1207.3083},
 primaryClass = {astro-ph.GA},
       adsurl = {https://ui.adsabs.harvard.edu/abs/2013MNRAS.429..159G}
}

@ARTICLE{Gomez2017,
       author = {{G{\'o}mez}, Facundo A. and {White}, Simon D.~M. and {Grand}, Robert J.~J. and {Marinacci}, Federico and {Springel}, Volker and {Pakmor}, R{\"u}diger},
        title = "{Warps and waves in the stellar discs of the Auriga cosmological simulations}",
      journal = {MNRAS},
         year = 2017,
        month = mar,
       volume = {465},
       number = {3},
        pages = {3446-3460},
          doi = {10.1093/mnras/stw2957},
archivePrefix = {arXiv},
       eprint = {1606.06295},
 primaryClass = {astro-ph.GA},
       adsurl = {https://ui.adsabs.harvard.edu/abs/2017MNRAS.465.3446G}
}

@ARTICLE{Gomez2016,
       author = {{G{\'o}mez}, Facundo A. and {White}, Simon D.~M. and {Marinacci}, Federico and {Slater}, Colin T. and {Grand}, Robert J.~J. and {Springel}, Volker and {Pakmor}, R{\"u}diger},
        title = "{A fully cosmological model of a Monoceros-like ring}",
      journal = {MNRAS},
         year = 2016,
        month = mar,
       volume = {456},
       number = {3},
        pages = {2779-2793},
          doi = {10.1093/mnras/stv2786},
archivePrefix = {arXiv},
       eprint = {1509.08459},
 primaryClass = {astro-ph.GA},
       adsurl = {https://ui.adsabs.harvard.edu/abs/2016MNRAS.456.2779G}
}

@ARTICLE{Bland-Hawthorn2021,
       author = {{Bland-Hawthorn}, Joss and {Tepper-Garc{\'\i}a}, Thor},
        title = "{Galactic seismology: the evolving 'phase spiral' after the Sagittarius dwarf impact}",
      journal = {\mnras},
         year = 2021,
        month = jul,
       volume = {504},
       number = {3},
        pages = {3168-3186},
          doi = {10.1093/mnras/stab704},
archivePrefix = {arXiv},
       eprint = {2009.02434},
 primaryClass = {astro-ph.GA},
       adsurl = {https://ui.adsabs.harvard.edu/abs/2021MNRAS.504.3168B}
}

@ARTICLE{Laporte2018,
       author = {{Laporte}, Chervin F.~P. and {Johnston}, Kathryn V. and {G{\'o}mez}, Facundo A. and {Garavito-Camargo}, Nicolas and {Besla}, Gurtina},
        title = "{The influence of Sagittarius and the Large Magellanic Cloud on the stellar disc of the Milky Way Galaxy}",
      journal = {MNRAS},
         year = 2018,
        month = nov,
       volume = {481},
       number = {1},
        pages = {286-306},
          doi = {10.1093/mnras/sty1574},
archivePrefix = {arXiv},
       eprint = {1710.02538},
 primaryClass = {astro-ph.GA},
       adsurl = {https://ui.adsabs.harvard.edu/abs/2018MNRAS.481..286L}
}

@ARTICLE{McMillan2022,
       author = {{McMillan}, Paul J. and {Petersson}, Jonathan and {Tepper-Garcia}, Thor and {Bland-Hawthorn}, Joss and {Antoja}, Teresa and {Chemin}, Laurent and {Figueras}, Francesca and {Khanna}, Shourya and {Kordopatis}, Georges and {Ramos}, Pau and {Romero-G{\'o}mez}, Merce and {Seabroke}, George},
        title = "{The disturbed outer Milky Way disc}",
      journal = {MNRAS},
         year = 2022,
        month = nov,
       volume = {516},
       number = {4},
        pages = {4988-5002},
          doi = {10.1093/mnras/stac2571},
archivePrefix = {arXiv},
       eprint = {2206.04059},
 primaryClass = {astro-ph.GA},
       adsurl = {https://ui.adsabs.harvard.edu/abs/2022MNRAS.516.4988M}
}

@ARTICLE{Momany2006,
       author = {{Momany}, Y. and {Zaggia}, S. and {Gilmore}, G. and {Piotto}, G. and {Carraro}, G. and {Bedin}, L.~R. and {de Angeli}, F.},
        title = "{Outer structure of the Galactic warp and flare: explaining the Canis Major over-density}",
      journal = {A\&A},
         year = 2006,
        month = may,
       volume = {451},
       number = {2},
        pages = {515-538},
          doi = {10.1051/0004-6361:20054081},
archivePrefix = {arXiv},
       eprint = {astro-ph/0603385},
 primaryClass = {astro-ph},
       adsurl = {https://ui.adsabs.harvard.edu/abs/2006A&A...451..515M}
}

@ARTICLE{Dhanush2024,
       author = {{Dhanush}, S.~R. and {Subramaniam}, A. and {Subramanian}, S.},
        title = "{A Comprehensive Kinematic Model of the Large Magellanic Cloud Disk from Star Clusters and Field Stars using Gaia DR3: Tracing the Disk Characteristics, Rotation, Bar, and Outliers}",
      journal = {ApJ},
         year = 2024,
        month = jun,
       volume = {968},
       number = {2},
          eid = {103},
        pages = {103},
          doi = {10.3847/1538-4357/ad4453},
archivePrefix = {arXiv},
       eprint = {2404.18658},
 primaryClass = {astro-ph.GA},
       adsurl = {https://ui.adsabs.harvard.edu/abs/2024ApJ...968..103D}
}

@ARTICLE{Pietrzynski2019,
       author = {{Pietrzy{\'n}ski}, G. and {Graczyk}, D. and {Gallenne}, A. and {Gieren}, W. and {Thompson}, I.~B. and {Pilecki}, B. and {Karczmarek}, P. and {G{\'o}rski}, M. and {Suchomska}, K. and {Taormina}, M. and {Zgirski}, B. and {Wielg{\'o}rski}, P. and {Ko{\l}aczkowski}, Z. and {Konorski}, P. and {Villanova}, S. and {Nardetto}, N. and {Kervella}, P. and {Bresolin}, F. and {Kudritzki}, R.~P. and {Storm}, J. and {Smolec}, R. and {Narloch}, W.},
        title = "{A distance to the Large Magellanic Cloud that is precise to one per cent}",
      journal = {\nat},
         year = 2019,
        month = mar,
       volume = {567},
       number = {7747},
        pages = {200-203},
          doi = {10.1038/s41586-019-0999-4},
archivePrefix = {arXiv},
       eprint = {1903.08096},
 primaryClass = {astro-ph.GA},
       adsurl = {https://ui.adsabs.harvard.edu/abs/2019Natur.567..200P}
}

@ARTICLE{Garver2026,
       author = {{Garver}, Bethany and {Nidever}, David and {Debattista}, Victor and {Deg}, Nathan},
        title = "{Modeling the recent interactions between the Magellanic Clouds and Milky Way}",
      journal = {arXiv e-prints},
         year = 2026,
        month = feb,
          eid = {arXiv:2602.05021},
        pages = {arXiv:2602.05021},
          doi = {10.48550/arXiv.2602.05021},
archivePrefix = {arXiv},
       eprint = {2602.05021},
 primaryClass = {astro-ph.GA},
       adsurl = {https://ui.adsabs.harvard.edu/abs/2026arXiv260205021G}
}

@ARTICLE{Oden2026,
       author = {{Oden}, Slater J. and {Nidever}, David L. and {Povick}, Joshua and {Massana}, Pol and {Choi}, Yumi and {van der Marel}, Roeland P. and {Cioni}, Maria-Rosa L. and {Sakowska}, Joanna and {Olsen}, Knut A.~G. and {Cullinane}, Lara and {Horta}, Danny and {Almeida}, Andres and {Carballo-Bello}, J.~A. and {Crnojevi{\'c}}, D. and {Ferguson}, P.~S. and {Mart{\'\i}nez-V{\'a}zquez}, C.~E. and {Medina}, G.~E. and {Mutlu-Pakdil}, B. and {Navabi}, M. and {Pace}, A.~B. and {Riley}, A.~H. and {Stringfellow}, Guy S. and {Vivas}, A.~K. and {James}, David J.},
        title = "{Warped and Hooked: Mapping the Magellanic Clouds in Three Dimensions Using Red Clump Stars}",
      journal = {\apj},
         year = 2026,
        month = may,
       volume = {1003},
       number = {1},
          eid = {9},
        pages = {9},
          doi = {10.3847/1538-4357/ae5c9e},
archivePrefix = {arXiv},
       eprint = {2512.04200},
 primaryClass = {astro-ph.GA},
       adsurl = {https://ui.adsabs.harvard.edu/abs/2026ApJ..1003....9O}
}

@ARTICLE{Wang2026,
       author = {{Wang}, Shuyu and {Brown}, Anthony G.~A. and {Debattista}, Victor P. and {Khachaturyants}, Tigran},
        title = "{Phase spirals induced by the gas warp}",
      journal = {\aap},
         year = 2026,
        month = jun,
       volume = {710},
          eid = {A120},
        pages = {A120},
          doi = {10.1051/0004-6361/202558812},
archivePrefix = {arXiv},
       eprint = {2604.06862},
 primaryClass = {astro-ph.GA},
       adsurl = {https://ui.adsabs.harvard.edu/abs/2026A&A...710A.120W}
}

@ARTICLE{Rathore2026b,
       author = {{Rathore}, H. and {Foote}, H. and {Besla}, G. and others},
        title = "{The Role of Dark Matter in Driving Disk Perturbations: Application to LMC-SMC Like Galaxy Interactions}",
	Year = {2026}
}

@ARTICLE{Nidever2026a,
       author = {{Nidever}, D. L. and {Massana}, P. and {Oden}, S. and others},
        title = "{DECam Local Volume Exploration Magellanic Clouds (DELVE-MC) Survey Overview}",
	Year = {2026a}
}

@ARTICLE{Ripepi2015,
       author = {{Ripepi}, V. and {Moretti}, M.~I. and {Marconi}, M. and {Clementini}, G. and {Cioni}, M.-R.~L. and {de Grijs}, R. and {Emerson}, J.~P. and {Groenewegen}, M.~A.~T. and {Ivanov}, V.~D. and {Muraveva}, T. and {Piatti}, A.~E. and {Subramanian}, S.},
        title = "{The VMC Survey - XIII. Type II Cepheids in the Large Magellanic Cloud}",
      journal = {\mnras},
         year = 2015,
        month = jan,
       volume = {446},
       number = {3},
        pages = {3034-3061},
          doi = {10.1093/mnras/stu2260},
archivePrefix = {arXiv},
       eprint = {1410.7817},
 primaryClass = {astro-ph.SR},
       adsurl = {https://ui.adsabs.harvard.edu/abs/2015MNRAS.446.3034R}
}

@ARTICLE{Widrow2012,
       author = {{Widrow}, Lawrence M. and {Gardner}, Susan and {Yanny}, Brian and {Dodelson}, Scott and {Chen}, Hsin-Yu},
        title = "{Galactoseismology: Discovery of Vertical Waves in the Galactic Disk}",
      journal = {\apjl},
         year = 2012,
        month = may,
       volume = {750},
       number = {2},
          eid = {L41},
        pages = {L41},
          doi = {10.1088/2041-8205/750/2/L41},
archivePrefix = {arXiv},
       eprint = {1203.6861},
 primaryClass = {astro-ph.GA},
       adsurl = {https://ui.adsabs.harvard.edu/abs/2012ApJ...750L..41W}
}

@ARTICLE{Yanny2013,
       author = {{Yanny}, Brian and {Gardner}, Susan},
        title = "{The Stellar Number Density Distribution in the Local Solar Neighborhood is North-South Asymmetric}",
      journal = {\apj},
         year = 2013,
        month = nov,
       volume = {777},
       number = {2},
          eid = {91},
        pages = {91},
          doi = {10.1088/0004-637X/777/2/91},
archivePrefix = {arXiv},
       eprint = {1309.2300},
 primaryClass = {astro-ph.GA},
       adsurl = {https://ui.adsabs.harvard.edu/abs/2013ApJ...777...91Y}
}

@ARTICLE{Xu2015,
       author = {{Xu}, Yan and {Newberg}, Heidi Jo and {Carlin}, Jeffrey L. and {Liu}, Chao and {Deng}, Licai and {Li}, Jing and {Sch{\"o}nrich}, Ralph and {Yanny}, Brian},
        title = "{Rings and Radial Waves in the Disk of the Milky Way}",
      journal = {\apj},
         year = 2015,
        month = mar,
       volume = {801},
       number = {2},
          eid = {105},
        pages = {105},
          doi = {10.1088/0004-637X/801/2/105},
archivePrefix = {arXiv},
       eprint = {1503.00257},
 primaryClass = {astro-ph.GA},
       adsurl = {https://ui.adsabs.harvard.edu/abs/2015ApJ...801..105X}
}

@ARTICLE{Antoja2018,
       author = {{Antoja}, T. and {Helmi}, A. and {Romero-G{\'o}mez}, M. and {Katz}, D. and {Babusiaux}, C. and {Drimmel}, R. and {Evans}, D.~W. and {Figueras}, F. and {Poggio}, E. and {Reyl{\'e}}, C. and {Robin}, A.~C. and {Seabroke}, G. and {Soubiran}, C.},
        title = "{A dynamically young and perturbed Milky Way disk}",
      journal = {\nat},
         year = 2018,
        month = sep,
       volume = {561},
       number = {7723},
        pages = {360-362},
          doi = {10.1038/s41586-018-0510-7},
archivePrefix = {arXiv},
       eprint = {1804.10196},
 primaryClass = {astro-ph.GA},
       adsurl = {https://ui.adsabs.harvard.edu/abs/2018Natur.561..360A}
}

@ARTICLE{Cullinane2022a,
       author = {{Cullinane}, L.~R. and {Mackey}, A.~D. and {Da Costa}, G.~S. and {Erkal}, D. and {Koposov}, S.~E. and {Belokurov}, V.},
        title = "{The Magellanic Edges Survey - II. Formation of the LMC's northern arm}",
      journal = {\mnras},
         year = 2022,
        month = feb,
       volume = {510},
       number = {1},
        pages = {445-468},
          doi = {10.1093/mnras/stab3350},
archivePrefix = {arXiv},
       eprint = {2106.03274},
 primaryClass = {astro-ph.GA},
       adsurl = {https://ui.adsabs.harvard.edu/abs/2022MNRAS.510..445C}
}

@ARTICLE{vandermarel2001a,
       author = {{van der Marel}, Roeland P. and {Cioni}, Maria-Rosa L.},
        title = "{Magellanic Cloud Structure from Near-Infrared Surveys. I. The Viewing Angles of the Large Magellanic Cloud}",
      journal = {\aj},
         year = 2001,
        month = oct,
       volume = {122},
       number = {4},
        pages = {1807-1826},
          doi = {10.1086/323099},
archivePrefix = {arXiv},
       eprint = {astro-ph/0105339},
 primaryClass = {astro-ph},
       adsurl = {https://ui.adsabs.harvard.edu/abs/2001AJ....122.1807V}
}

@ARTICLE{vandermarel2002,
       author = {{van der Marel}, Roeland P. and {Alves}, David R. and {Hardy}, Eduardo and {Suntzeff}, Nicholas B.},
        title = "{New Understanding of Large Magellanic Cloud Structure, Dynamics, and Orbit from Carbon Star Kinematics}",
      journal = {\aj},
         year = 2002,
        month = nov,
       volume = {124},
       number = {5},
        pages = {2639-2663},
          doi = {10.1086/343775},
archivePrefix = {arXiv},
       eprint = {astro-ph/0205161},
 primaryClass = {astro-ph},
       adsurl = {https://ui.adsabs.harvard.edu/abs/2002AJ....124.2639V}
}

@ARTICLE{JimenezArranz2023,
       author = {{Jim{\'e}nez-Arranz}, {\'O}. and {Romero-G{\'o}mez}, M. and {Luri}, X. and {McMillan}, P.~J. and {Antoja}, T. and {Chemin}, L. and {Roca-F{\`a}brega}, S. and {Masana}, E. and {Muros}, A.},
        title = "{Kinematic analysis of the Large Magellanic Cloud using Gaia DR3}",
      journal = {\aap},
         year = 2023,
        month = jan,
       volume = {669},
          eid = {A91},
        pages = {A91},
          doi = {10.1051/0004-6361/202244601},
archivePrefix = {arXiv},
       eprint = {2210.01728},
 primaryClass = {astro-ph.GA},
       adsurl = {https://ui.adsabs.harvard.edu/abs/2023A&A...669A..91J}
}

@ARTICLE{JimenezArranz2024a,
       author = {{Jim{\'e}nez-Arranz}, {\'O}. and {Chemin}, L. and {Romero-G{\'o}mez}, M. and {Luri}, X. and {Adamczyk}, P. and {Castro-Ginard}, A. and {Roca-F{\`a}brega}, S. and {McMillan}, P.~J. and {Cioni}, M.-R.~L.},
        title = "{The bar pattern speed of the Large Magellanic Cloud}",
      journal = {\aap},
         year = 2024,
        month = mar,
       volume = {683},
          eid = {A102},
        pages = {A102},
          doi = {10.1051/0004-6361/202347266},
archivePrefix = {arXiv},
       eprint = {2312.11192},
 primaryClass = {astro-ph.GA},
       adsurl = {https://ui.adsabs.harvard.edu/abs/2024A&A...683A.102J}
}

@ARTICLE{JimenezArranz2025,
       author = {{Jim{\'e}nez-Arranz}, {\'O}. and {Horta}, D. and {van der Marel}, R.~P. and {Nidever}, D. and {Laporte}, C.~F.~P. and {Patel}, E. and {Rix}, H. -W.},
        title = "{Vertical structure and kinematics of the LMC disc from SDSS/Gaia}",
      journal = {\aap},
         year = 2025,
        month = jun,
       volume = {698},
          eid = {A88},
        pages = {A88},
          doi = {10.1051/0004-6361/202553705},
archivePrefix = {arXiv},
       eprint = {2501.04616},
 primaryClass = {astro-ph.GA},
       adsurl = {https://ui.adsabs.harvard.edu/abs/2025A&A...698A..88J}
}

@ARTICLE{Besla2012,
       author = {{Besla}, Gurtina and {Kallivayalil}, Nitya and {Hernquist}, Lars and {van der Marel}, Roeland P. and {Cox}, T.~J. and {Kere{\v{s}}}, Du{\v{s}}an},
        title = "{The role of dwarf galaxy interactions in shaping the Magellanic System and implications for Magellanic Irregulars}",
      journal = {\mnras},
         year = 2012,
        month = apr,
       volume = {421},
       number = {3},
        pages = {2109-2138},
          doi = {10.1111/j.1365-2966.2012.20466.x},
archivePrefix = {arXiv},
       eprint = {1201.1299},
 primaryClass = {astro-ph.GA},
       adsurl = {https://ui.adsabs.harvard.edu/abs/2012MNRAS.421.2109B}
}

@ARTICLE{Zivick2018,
       author = {{Zivick}, Paul and {Kallivayalil}, Nitya and {van der Marel}, Roeland P. and {Besla}, Gurtina and {Linden}, Sean T. and {Koz{\l}owski}, Szymon and {Fritz}, Tobias K. and {Kochanek}, C.~S. and {Anderson}, J. and {Sohn}, Sangmo Tony and {Geha}, Marla C. and {Alcock}, Charles R.},
        title = "{The Proper Motion Field of the Small Magellanic Cloud: Kinematic Evidence for Its Tidal Disruption}",
      journal = {\apj},
         year = 2018,
        month = sep,
       volume = {864},
       number = {1},
          eid = {55},
        pages = {55},
          doi = {10.3847/1538-4357/aad4b0},
archivePrefix = {arXiv},
       eprint = {1804.04110},
 primaryClass = {astro-ph.GA},
       adsurl = {https://ui.adsabs.harvard.edu/abs/2018ApJ...864...55Z}
}

@ARTICLE{Zivick2019,
       author = {{Zivick}, Paul and {Kallivayalil}, Nitya and {Besla}, Gurtina and {Sohn}, Sangmo Tony and {van der Marel}, Roeland P. and {del Pino}, Andr{\'e}s and {Linden}, Sean T. and {Fritz}, Tobias K. and {Anderson}, J.},
        title = "{The Proper-motion Field along the Magellanic Bridge: A New Probe of the LMC-SMC Interaction}",
      journal = {\apj},
         year = 2019,
        month = mar,
       volume = {874},
       number = {1},
          eid = {78},
        pages = {78},
          doi = {10.3847/1538-4357/ab0554},
archivePrefix = {arXiv},
       eprint = {1811.09318},
 primaryClass = {astro-ph.GA},
       adsurl = {https://ui.adsabs.harvard.edu/abs/2019ApJ...874...78Z}
}

@ARTICLE{Zhao2000,
       author = {{Zhao}, HongSheng and {Evans}, N. Wyn},
        title = "{The So-called ``Bar'' in the Large Magellanic Cloud}",
      journal = {\apjl},
         year = 2000,
        month = dec,
       volume = {545},
       number = {1},
        pages = {L35-L38},
          doi = {10.1086/317324},
archivePrefix = {arXiv},
       eprint = {astro-ph/0009155},
 primaryClass = {astro-ph},
       adsurl = {https://ui.adsabs.harvard.edu/abs/2000ApJ...545L..35Z}
}

@ARTICLE{Rathore2025,
       author = {{Rathore}, Himansh and {Choi}, Yumi and {Olsen}, Knut A.~G. and {Besla}, Gurtina},
        title = "{Precise Measurements of the LMC Bar's Geometry with Gaia DR3 and a Novel Solution to Crowding-induced Incompleteness in Star Counting}",
      journal = {\apj},
         year = 2025,
        month = jan,
       volume = {978},
       number = {1},
          eid = {55},
        pages = {55},
          doi = {10.3847/1538-4357/ad93ae},
archivePrefix = {arXiv},
       eprint = {2410.18182},
 primaryClass = {astro-ph.GA},
       adsurl = {https://ui.adsabs.harvard.edu/abs/2025ApJ...978...55R}
}

@ARTICLE{Rathore2026,
       author = {{Rathore}, Himansh and {Besla}, Gurtina and {van der Marel}, Roeland P. and {Kallivayalil}, Nitya},
        title = "{A Galactic Transformation{\textemdash}Understanding the SMC's Structural and Kinematic Disequilibrium}",
      journal = {\apj},
         year = 2026,
        month = mar,
       volume = {1000},
       number = {1},
          eid = {50},
        pages = {50},
          doi = {10.3847/1538-4357/ae4507},
archivePrefix = {arXiv},
       eprint = {2512.06075},
 primaryClass = {astro-ph.GA},
       adsurl = {https://ui.adsabs.harvard.edu/abs/2026ApJ..1000...50R}
}

@ARTICLE{elyoussoufi2021,
       author = {{El Youssoufi}, Dalal and {Cioni}, Maria-Rosa L. and {Bell}, Cameron P.~M. and {de Grijs}, Richard and {Groenewegen}, Martin A.~T. and {Ivanov}, Valentin D. and {Matijev{\u{i}}c}, Gal and {Niederhofer}, Florian and {Oliveira}, Joana M. and {Ripepi}, Vincenzo and {Schmidt}, Thomas and {Subramanian}, Smitha and {Sun}, Ning-Chen and {van Loon}, Jacco Th},
        title = "{Stellar substructures in the periphery of the Magellanic Clouds with the VISTA hemisphere survey from the red clump and other tracers}",
      journal = {\mnras},
         year = 2021,
        month = aug,
       volume = {505},
       number = {2},
        pages = {2020-2038},
          doi = {10.1093/mnras/stab1075},
archivePrefix = {arXiv},
       eprint = {2104.08054},
 primaryClass = {astro-ph.GA},
       adsurl = {https://ui.adsabs.harvard.edu/abs/2021MNRAS.505.2020E}
}

@ARTICLE{Mackey2016,
       author = {{Mackey}, A.~D. and {Koposov}, S.~E. and {Erkal}, D. and {Belokurov}, V. and {Da Costa}, G.~S. and {G{\'o}mez}, F.~A.},
        title = "{A 10 kpc stellar substructure at the edge of the Large Magellanic Cloud: perturbed outer disc or evidence for tidal stripping?}",
      journal = {\mnras},
         year = 2016,
        month = jun,
       volume = {459},
       number = {1},
        pages = {239-255},
          doi = {10.1093/mnras/stw497},
archivePrefix = {arXiv},
       eprint = {1508.01356},
 primaryClass = {astro-ph.GA},
       adsurl = {https://ui.adsabs.harvard.edu/abs/2016MNRAS.459..239M}
}

@ARTICLE{Mackey2017,
       author = {{Mackey}, A.~D. and {Koposov}, S.~E. and {Da Costa}, G.~S. and {Belokurov}, V. and {Erkal}, D. and {Fraternali}, F. and {McClure-Griffiths}, N.~M. and {Fraser}, M.},
        title = "{Structured star formation in the Magellanic inter-Cloud region}",
      journal = {\mnras},
         year = 2017,
        month = dec,
       volume = {472},
       number = {3},
        pages = {2975-2989},
          doi = {10.1093/mnras/stx2035},
archivePrefix = {arXiv},
       eprint = {1708.04363},
 primaryClass = {astro-ph.GA},
       adsurl = {https://ui.adsabs.harvard.edu/abs/2017MNRAS.472.2975M}
}

@ARTICLE{GaiaCollaboration2016,
       author = {{Gaia Collaboration} and {Prusti}, T. and {de Bruijne}, J.~H.~J. and {Brown}, A.~G.~A. and {Vallenari}, A. and {Babusiaux}, C. and {Bailer-Jones}, C.~A.~L. and {Bastian}, U. and {Biermann}, M. and {Evans}, D.~W. and {Eyer}, L. and {Jansen}, F. and {Jordi}, C. and {Klioner}, S.~A. and {Lammers}, U. and {Lindegren}, L. and {Luri}, X. and {Mignard}, F. and {Milligan}, D.~J. and {Panem}, C. and {Poinsignon}, V. and {Pourbaix}, D. and {Randich}, S. and {Sarri}, G. and {Sartoretti}, P. and {Siddiqui}, H.~I. and {Soubiran}, C. and {Valette}, V. and {van Leeuwen}, F. and {Walton}, N.~A. and {Aerts}, C. and {Arenou}, F. and {Cropper}, M. and {Drimmel}, R. and {H{\o}g}, E. and {Katz}, D. and {Lattanzi}, M.~G. and {O'Mullane}, W. and {Grebel}, E.~K. and {Holland}, A.~D. and {Huc}, C. and {Passot}, X. and {Bramante}, L. and {Cacciari}, C. and {Casta{\~n}eda}, J. and {Chaoul}, L. and {Cheek}, N. and {De Angeli}, F. and {Fabricius}, C. and {Guerra}, R. and {Hern{\'a}ndez}, J. and {Jean-Antoine-Piccolo}, A. and {Masana}, E. and {Messineo}, R. and {Mowlavi}, N. and {Nienartowicz}, K. and {Ord{\'o}{\~n}ez-Blanco}, D. and {Panuzzo}, P. and {Portell}, J. and {Richards}, P.~J. and {Riello}, M. and {Seabroke}, G.~M. and {Tanga}, P. and {Th{\'e}venin}, F. and {Torra}, J. and {Els}, S.~G. and {Gracia-Abril}, G. and {Comoretto}, G. and {Garcia-Reinaldos}, M. and {Lock}, T. and {Mercier}, E. and {Altmann}, M. and {Andrae}, R. and {Astraatmadja}, T.~L. and {Bellas-Velidis}, I. and {Benson}, K. and {Berthier}, J. and {Blomme}, R. and {Busso}, G. and {Carry}, B. and {Cellino}, A. and {Clementini}, G. and {Cowell}, S. and {Creevey}, O. and {Cuypers}, J. and {Davidson}, M. and {De Ridder}, J. and {de Torres}, A. and {Delchambre}, L. and {Dell'Oro}, A. and {Ducourant}, C. and {Fr{\'e}mat}, Y. and {Garc{\'\i}a-Torres}, M. and {Gosset}, E. and {Halbwachs}, J. -L. and {Hambly}, N.~C. and {Harrison}, D.~L. and {Hauser}, M. and {Hestroffer}, D. and {Hodgkin}, S.~T. and {Huckle}, H.~E. and {Hutton}, A. and {Jasniewicz}, G. and {Jordan}, S. and {Kontizas}, M. and {Korn}, A.~J. and {Lanzafame}, A.~C. and {Manteiga}, M. and {Moitinho}, A. and {Muinonen}, K. and {Osinde}, J. and {Pancino}, E. and {Pauwels}, T. and {Petit}, J. -M. and {Recio-Blanco}, A. and {Robin}, A.~C. and {Sarro}, L.~M. and {Siopis}, C. and {Smith}, M. and {Smith}, K.~W. and {Sozzetti}, A. and {Thuillot}, W. and {van Reeven}, W. and {Viala}, Y. and {Abbas}, U. and {Abreu Aramburu}, A. and {Accart}, S. and {Aguado}, J.~J. and {Allan}, P.~M. and {Allasia}, W. and {Altavilla}, G. and {{\'A}lvarez}, M.~A. and {Alves}, J. and {Anderson}, R.~I. and {Andrei}, A.~H. and {Anglada Varela}, E. and {Antiche}, E. and {Antoja}, T. and {Ant{\'o}n}, S. and {Arcay}, B. and {Atzei}, A. and {Ayache}, L. and {Bach}, N. and {Baker}, S.~G. and {Balaguer-N{\'u}{\~n}ez}, L. and {Barache}, C. and {Barata}, C. and {Barbier}, A. and {Barblan}, F. and {Baroni}, M. and {Barrado y Navascu{\'e}s}, D. and {Barros}, M. and {Barstow}, M.~A. and {Becciani}, U. and {Bellazzini}, M. and {Bellei}, G. and {Bello Garc{\'\i}a}, A. and {Belokurov}, V. and {Bendjoya}, P. and {Berihuete}, A. and {Bianchi}, L. and {Bienaym{\'e}}, O. and {Billebaud}, F. and {Blagorodnova}, N. and {Blanco-Cuaresma}, S. and {Boch}, T. and {Bombrun}, A. and {Borrachero}, R. and {Bouquillon}, S. and {Bourda}, G. and {Bouy}, H. and {Bragaglia}, A. and {Breddels}, M.~A. and {Brouillet}, N. and {Br{\"u}semeister}, T. and {Bucciarelli}, B. and {Budnik}, F. and {Burgess}, P. and {Burgon}, R. and {Burlacu}, A. and {Busonero}, D. and {Buzzi}, R. and {Caffau}, E. and {Cambras}, J. and {Campbell}, H. and {Cancelliere}, R. and {Cantat-Gaudin}, T. and {Carlucci}, T. and {Carrasco}, J.~M. and {Castellani}, M. and {Charlot}, P. and {Charnas}, J. and {Charvet}, P. and {Chassat}, F. and {Chiavassa}, A. and {Clotet}, M. and {Cocozza}, G. and {Collins}, R.~S. and {Collins}, P. and {Costigan}, G.},
        title = "{The Gaia mission}",
      journal = {\aap},
         year = 2016,
        month = nov,
       volume = {595},
          eid = {A1},
        pages = {A1},
          doi = {10.1051/0004-6361/201629272},
archivePrefix = {arXiv},
       eprint = {1609.04153},
 primaryClass = {astro-ph.IM},
       adsurl = {https://ui.adsabs.harvard.edu/abs/2016A&A...595A...1G}
}

@ARTICLE{GaiaCollaboration2021a,
       author = {{Gaia Collaboration} and {Brown}, A.~G.~A. and {Vallenari}, A. and {Prusti}, T. and {de Bruijne}, J.~H.~J. and {Babusiaux}, C. and {Biermann}, M. and {Creevey}, O.~L. and {Evans}, D.~W. and {Eyer}, L. and {Hutton}, A. and {Jansen}, F. and {Jordi}, C. and {Klioner}, S.~A. and {Lammers}, U. and {Lindegren}, L. and {Luri}, X. and {Mignard}, F. and {Panem}, C. and {Pourbaix}, D. and {Randich}, S. and {Sartoretti}, P. and {Soubiran}, C. and {Walton}, N.~A. and {Arenou}, F. and {Bailer-Jones}, C.~A.~L. and {Bastian}, U. and {Cropper}, M. and {Drimmel}, R. and {Katz}, D. and {Lattanzi}, M.~G. and {van Leeuwen}, F. and {Bakker}, J. and {Cacciari}, C. and {Casta{\~n}eda}, J. and {De Angeli}, F. and {Ducourant}, C. and {Fabricius}, C. and {Fouesneau}, M. and {Fr{\'e}mat}, Y. and {Guerra}, R. and {Guerrier}, A. and {Guiraud}, J. and {Jean-Antoine Piccolo}, A. and {Masana}, E. and {Messineo}, R. and {Mowlavi}, N. and {Nicolas}, C. and {Nienartowicz}, K. and {Pailler}, F. and {Panuzzo}, P. and {Riclet}, F. and {Roux}, W. and {Seabroke}, G.~M. and {Sordo}, R. and {Tanga}, P. and {Th{\'e}venin}, F. and {Gracia-Abril}, G. and {Portell}, J. and {Teyssier}, D. and {Altmann}, M. and {Andrae}, R. and {Bellas-Velidis}, I. and {Benson}, K. and {Berthier}, J. and {Blomme}, R. and {Brugaletta}, E. and {Burgess}, P.~W. and {Busso}, G. and {Carry}, B. and {Cellino}, A. and {Cheek}, N. and {Clementini}, G. and {Damerdji}, Y. and {Davidson}, M. and {Delchambre}, L. and {Dell'Oro}, A. and {Fern{\'a}ndez-Hern{\'a}ndez}, J. and {Galluccio}, L. and {Garc{\'\i}a-Lario}, P. and {Garcia-Reinaldos}, M. and {Gonz{\'a}lez-N{\'u}{\~n}ez}, J. and {Gosset}, E. and {Haigron}, R. and {Halbwachs}, J. -L. and {Hambly}, N.~C. and {Harrison}, D.~L. and {Hatzidimitriou}, D. and {Heiter}, U. and {Hern{\'a}ndez}, J. and {Hestroffer}, D. and {Hodgkin}, S.~T. and {Holl}, B. and {Jan{\ss}en}, K. and {Jevardat de Fombelle}, G. and {Jordan}, S. and {Krone-Martins}, A. and {Lanzafame}, A.~C. and {L{\"o}ffler}, W. and {Lorca}, A. and {Manteiga}, M. and {Marchal}, O. and {Marrese}, P.~M. and {Moitinho}, A. and {Mora}, A. and {Muinonen}, K. and {Osborne}, P. and {Pancino}, E. and {Pauwels}, T. and {Petit}, J. -M. and {Recio-Blanco}, A. and {Richards}, P.~J. and {Riello}, M. and {Rimoldini}, L. and {Robin}, A.~C. and {Roegiers}, T. and {Rybizki}, J. and {Sarro}, L.~M. and {Siopis}, C. and {Smith}, M. and {Sozzetti}, A. and {Ulla}, A. and {Utrilla}, E. and {van Leeuwen}, M. and {van Reeven}, W. and {Abbas}, U. and {Abreu Aramburu}, A. and {Accart}, S. and {Aerts}, C. and {Aguado}, J.~J. and {Ajaj}, M. and {Altavilla}, G. and {{\'A}lvarez}, M.~A. and {{\'A}lvarez Cid-Fuentes}, J. and {Alves}, J. and {Anderson}, R.~I. and {Anglada Varela}, E. and {Antoja}, T. and {Audard}, M. and {Baines}, D. and {Baker}, S.~G. and {Balaguer-N{\'u}{\~n}ez}, L. and {Balbinot}, E. and {Balog}, Z. and {Barache}, C. and {Barbato}, D. and {Barros}, M. and {Barstow}, M.~A. and {Bartolom{\'e}}, S. and {Bassilana}, J. -L. and {Bauchet}, N. and {Baudesson-Stella}, A. and {Becciani}, U. and {Bellazzini}, M. and {Bernet}, M. and {Bertone}, S. and {Bianchi}, L. and {Blanco-Cuaresma}, S. and {Boch}, T. and {Bombrun}, A. and {Bossini}, D. and {Bouquillon}, S. and {Bragaglia}, A. and {Bramante}, L. and {Breedt}, E. and {Bressan}, A. and {Brouillet}, N. and {Bucciarelli}, B. and {Burlacu}, A. and {Busonero}, D. and {Butkevich}, A.~G. and {Buzzi}, R. and {Caffau}, E. and {Cancelliere}, R. and {C{\'a}novas}, H. and {Cantat-Gaudin}, T. and {Carballo}, R. and {Carlucci}, T. and {Carnerero}, M.~I. and {Carrasco}, J.~M. and {Casamiquela}, L. and {Castellani}, M. and {Castro-Ginard}, A. and {Castro Sampol}, P. and {Chaoul}, L. and {Charlot}, P. and {Chemin}, L. and {Chiavassa}, A. and {Cioni}, M. -R.~L. and {Comoretto}, G. and {Cooper}, W.~J. and {Cornez}, T. and {Cowell}, S. and {Crifo}, F. and {Crosta}, M. and {Crowley}, C. and {Dafonte}, C. and {Dapergolas}, A. and {David}, M. and {David}, P.},
        title = "{Gaia Early Data Release 3. Summary of the contents and survey properties}",
      journal = {\aap},
         year = 2021,
        month = may,
       volume = {649},
          eid = {A1},
        pages = {A1},
          doi = {10.1051/0004-6361/202039657},
archivePrefix = {arXiv},
       eprint = {2012.01533},
 primaryClass = {astro-ph.GA},
       adsurl = {https://ui.adsabs.harvard.edu/abs/2021A&A...649A...1G}
}

@ARTICLE{Cheng2022,
       author = {{Cheng}, Xinlun and {Choi}, Yumi and {Olsen}, Knut and {Nidever}, David L. and {Majewski}, Steven R. and {Monachesi}, Antonela and {Besla}, Gurtina and {Mu{\~n}oz Gonzalez}, C{\'e}sar and {Anguiano}, Borja and {Almeida}, Andres and {Mu{\~n}oz}, Ricardo R. and {Lane}, Richard R. and {Nitschelm}, Christian},
        title = "{Kinematical Analysis of Substructure in the Southern Periphery of the Large Magellanic Cloud}",
      journal = {\apj},
         year = 2022,
        month = mar,
       volume = {928},
       number = {1},
          eid = {95},
        pages = {95},
          doi = {10.3847/1538-4357/ac5621},
archivePrefix = {arXiv},
       eprint = {2202.12789},
 primaryClass = {astro-ph.GA},
       adsurl = {https://ui.adsabs.harvard.edu/abs/2022ApJ...928...95C}
}

@ARTICLE{Harris2009,
       author = {{Harris}, Jason and {Zaritsky}, Dennis},
        title = "{The Star Formation History of the Large Magellanic Cloud}",
      journal = {\aj},
         year = 2009,
        month = nov,
       volume = {138},
       number = {5},
        pages = {1243-1260},
          doi = {10.1088/0004-6256/138/5/1243},
archivePrefix = {arXiv},
       eprint = {0908.1422},
 primaryClass = {astro-ph.CO},
       adsurl = {https://ui.adsabs.harvard.edu/abs/2009AJ....138.1243H}
}

@ARTICLE{JacyszynDobrzeniecka2016,
       author = {{Jacyszyn-Dobrzeniecka}, A.~M. and {Skowron}, D.~M. and {Mr{\'o}z}, P. and {Skowron}, J. and {Soszy{\'n}ski}, I. and {Udalski}, A. and {Pietrukowicz}, P. and {Koz{\l}owski}, S. and {Wyrzykowski}, {\L}. and {Poleski}, R. and {Pawlak}, M. and {Szyma{\'n}ski}, M.~K. and {Ulaczyk}, K.},
        title = "{OGLE-ing the Magellanic System: Three-Dimensional Structure of the Clouds and the Bridge Using Classical Cepheids}",
      journal = {\actaa},
         year = 2016,
        month = jun,
       volume = {66},
       number = {2},
        pages = {149-196},
          doi = {10.48550/arXiv.1602.09141},
archivePrefix = {arXiv},
       eprint = {1602.09141},
 primaryClass = {astro-ph.GA},
       adsurl = {https://ui.adsabs.harvard.edu/abs/2016AcA....66..149J}
}

@ARTICLE{Bitsakis2018,
       author = {{Bitsakis}, Theodoros and {Gonz{\'a}lez-L{\'o}pezlira}, R.~A. and {Bonfini}, P. and {Bruzual}, G. and {Maravelias}, G. and {Zaritsky}, D. and {Charlot}, S. and {Ram{\'\i}rez-Siordia}, V.~H.},
        title = "{The Distribution and Ages of Star Clusters in the Small Magellanic Cloud: Constraints on the Interaction History of the Magellanic Clouds}",
      journal = {\apj},
         year = 2018,
        month = feb,
       volume = {853},
       number = {2},
          eid = {104},
        pages = {104},
          doi = {10.3847/1538-4357/aaa244},
archivePrefix = {arXiv},
       eprint = {1712.04974},
 primaryClass = {astro-ph.GA},
       adsurl = {https://ui.adsabs.harvard.edu/abs/2018ApJ...853..104B}
}

@ARTICLE{Besla2016,
       author = {{Besla}, Gurtina and {Mart{\'\i}nez-Delgado}, David and {van der Marel}, Roeland P. and {Beletsky}, Yuri and {Seibert}, Mark and {Schlafly}, Edward F. and {Grebel}, Eva K. and {Neyer}, Fabian},
        title = "{Low Surface Brightness Imaging of the Magellanic System: Imprints of Tidal Interactions between the Clouds in the Stellar Periphery}",
      journal = {\apj},
         year = 2016,
        month = jul,
       volume = {825},
       number = {1},
          eid = {20},
        pages = {20},
          doi = {10.3847/0004-637X/825/1/20},
archivePrefix = {arXiv},
       eprint = {1602.04222},
 primaryClass = {astro-ph.GA},
       adsurl = {https://ui.adsabs.harvard.edu/abs/2016ApJ...825...20B}
}

@ARTICLE{Belokurov2019,
       author = {{Belokurov}, Vasily A. and {Erkal}, Denis},
        title = "{Clouds in arms}",
      journal = {\mnras},
         year = 2019,
        month = jan,
       volume = {482},
       number = {1},
        pages = {L9-L13},
          doi = {10.1093/mnrasl/sly178},
archivePrefix = {arXiv},
       eprint = {1808.00462},
 primaryClass = {astro-ph.GA},
       adsurl = {https://ui.adsabs.harvard.edu/abs/2019MNRAS.482L...9B}
}

@ARTICLE{devaucouleurs1972,
       author = {{de Vaucouleurs}, G. and {Freeman}, K.~C.},
        title = "{Structure and dynamics of barred spiral galaxies, in particular of the Magellanic type}",
      journal = {Vistas in Astronomy},
         year = 1972,
        month = jan,
       volume = {14},
       number = {1},
        pages = {163-294},
          doi = {10.1016/0083-6656(72)90026-8},
       adsurl = {https://ui.adsabs.harvard.edu/abs/1972VA.....14..163D}
}

@ARTICLE{Madore2025,
       author = {{Madore}, Barry F. and {Freedman}, Wendy L. and {Hoyt}, Taylor and {Jang}, In Sung and {Lee}, Abigail J. and {Owens}, Kayla A.},
        title = "{I-band Asymptotic Giant Branch Stars. I. Exploring a New Standard Candle for the Extragalactic Distance Scale}",
      journal = {\aj},
         year = 2025,
        month = mar,
       volume = {169},
       number = {3},
          eid = {162},
        pages = {162},
          doi = {10.3847/1538-3881/ada3d4},
       adsurl = {https://ui.adsabs.harvard.edu/abs/2025AJ....169..162M}
}

@ARTICLE{Groenewegen2019,
       author = {{Groenewegen}, M.~A.~T. and {Cioni}, M.-R.~L. and {Girardi}, L. and {de Grijs}, R. and {Ivanov}, V.~D. and {Marconi}, M. and {Muraveva}, T. and {Ripepi}, V. and {van Loon}, J. Th.},
        title = "{The VMC Survey. XXXIII. The tip of the red giant branch in the Magellanic Clouds}",
      journal = {\aap},
         year = 2019,
        month = feb,
       volume = {622},
          eid = {A63},
        pages = {A63},
          doi = {10.1051/0004-6361/201833904},
archivePrefix = {arXiv},
       eprint = {1812.02581},
 primaryClass = {astro-ph.SR},
       adsurl = {https://ui.adsabs.harvard.edu/abs/2019A&A...622A..63G}
}

@ARTICLE{Borissova2009,
       author = {{Borissova}, J. and {Rejkuba}, M. and {Minniti}, D. and {Catelan}, M. and {Ivanov}, V.~D.},
        title = "{Properties of RR Lyrae stars in the inner regions of the Large Magellanic Cloud. III. Near-infrared study}",
      journal = {\aap},
         year = 2009,
        month = aug,
       volume = {502},
       number = {2},
        pages = {505-514},
          doi = {10.1051/0004-6361/200811465},
       adsurl = {https://ui.adsabs.harvard.edu/abs/2009A&A...502..505B}
}

@ARTICLE{Choi2018a,
       author = {{Choi}, Yumi and {Nidever}, David L. and {Olsen}, Knut and {Blum}, Robert D. and {Besla}, Gurtina and {Zaritsky}, Dennis and {van der Marel}, Roeland P. and {Bell}, Eric F. and {Gallart}, Carme and {Cioni}, Maria-Rosa L. and {Johnson}, L. Clifton and {Vivas}, A. Katherina and {Saha}, Abhijit and {de Boer}, Thomas J.~L. and {No{\"e}l}, Noelia E.~D. and {Monachesi}, Antonela and {Massana}, Pol and {Conn}, Blair C. and {Martinez-Delgado}, David and {Mu{\~n}oz}, Ricardo R. and {Stringfellow}, Guy S.},
        title = "{SMASHing the LMC: A Tidally Induced Warp in the Outer LMC and a Large-scale Reddening Map}",
      journal = {\apj},
         year = 2018,
        month = oct,
       volume = {866},
       number = {2},
          eid = {90},
        pages = {90},
          doi = {10.3847/1538-4357/aae083},
archivePrefix = {arXiv},
       eprint = {1804.07765},
 primaryClass = {astro-ph.GA},
       adsurl = {https://ui.adsabs.harvard.edu/abs/2018ApJ...866...90C}
}

@ARTICLE{Choi2018b,
       author = {{Choi}, Yumi and {Nidever}, David L. and {Olsen}, Knut and {Besla}, Gurtina and {Blum}, Robert D. and {Zaritsky}, Dennis and {Cioni}, Maria-Rosa L. and {van der Marel}, Roeland P. and {Bell}, Eric F. and {Johnson}, L. Clifton and {Vivas}, A. Katherina and {Walker}, Alistair R. and {de Boer}, Thomas J.~L. and {No{\"e}l}, Noelia E.~D. and {Monachesi}, Antonela and {Gallart}, Carme and {Monelli}, Matteo and {Stringfellow}, Guy S. and {Massana}, Pol and {Martinez-Delgado}, David and {Mu{\~n}oz}, Ricardo R.},
        title = "{SMASHing the LMC: Mapping a Ring-like Stellar Overdensity in the LMC Disk}",
      journal = {\apj},
         year = 2018,
        month = dec,
       volume = {869},
       number = {2},
          eid = {125},
        pages = {125},
          doi = {10.3847/1538-4357/aaed1f},
archivePrefix = {arXiv},
       eprint = {1805.00481},
 primaryClass = {astro-ph.GA},
       adsurl = {https://ui.adsabs.harvard.edu/abs/2018ApJ...869..125C}
}

@ARTICLE{Saroon2022,
       author = {{Saroon}, S. and {Subramanian}, S.},
        title = "{Shape of the outer stellar warp in the Large Magellanic Cloud disk}",
      journal = {\aap},
         year = 2022,
        month = oct,
       volume = {666},
          eid = {A103},
        pages = {A103},
          doi = {10.1051/0004-6361/202141435},
archivePrefix = {arXiv},
       eprint = {2207.13269},
 primaryClass = {astro-ph.GA},
       adsurl = {https://ui.adsabs.harvard.edu/abs/2022A&A...666A.103S}
}

@ARTICLE{Olsen2002,
       author = {{Olsen}, K.~A.~G. and {Salyk}, C.},
        title = "{A Warp in the Large Magellanic Cloud Disk?}",
      journal = {\aj},
         year = 2002,
        month = oct,
       volume = {124},
       number = {4},
        pages = {2045-2053},
          doi = {10.1086/342739},
       adsurl = {https://ui.adsabs.harvard.edu/abs/2002AJ....124.2045O}
}

@ARTICLE{Choi2022,
       author = {{Choi}, Yumi and {Olsen}, Knut A.~G. and {Besla}, Gurtina and {van der Marel}, Roeland P. and {Zivick}, Paul and {Kallivayalil}, Nitya and {Nidever}, David L.},
        title = "{The Recent LMC-SMC Collision: Timing and Impact Parameter Constraints from Comparison of Gaia LMC Disk Kinematics and N-body Simulations}",
      journal = {\apj},
         year = 2022,
        month = mar,
       volume = {927},
       number = {2},
          eid = {153},
        pages = {153},
          doi = {10.3847/1538-4357/ac4e90},
archivePrefix = {arXiv},
       eprint = {2201.04648},
 primaryClass = {astro-ph.GA},
       adsurl = {https://ui.adsabs.harvard.edu/abs/2022ApJ...927..153C}
}

@ARTICLE{Skowron2021,
       author = {{Skowron}, D.~M. and {Skowron}, J. and {Udalski}, A. and {Szyma{\'n}ski}, M.~K. and {Soszy{\'n}ski}, I. and {Wyrzykowski}, {\L}. and {Ulaczyk}, K. and {Poleski}, R. and {Koz{\l}owski}, S. and {Pietrukowicz}, P. and {Mr{\'o}z}, P. and {Rybicki}, K. and {Iwanek}, P. and {Wrona}, M. and {Gromadzki}, M.},
        title = "{OGLE-ing the Magellanic System: Optical Reddening Maps of the Large and Small Magellanic Clouds from Red Clump Stars}",
      journal = {\apjs},
         year = 2021,
        month = feb,
       volume = {252},
       number = {2},
          eid = {23},
        pages = {23},
          doi = {10.3847/1538-4365/abcb81},
archivePrefix = {arXiv},
       eprint = {2006.02448},
 primaryClass = {astro-ph.SR},
       adsurl = {https://ui.adsabs.harvard.edu/abs/2021ApJS..252...23S}
}

@ARTICLE{Schlegel1998,
       author = {{Schlegel}, David J. and {Finkbeiner}, Douglas P. and {Davis}, Marc},
        title = "{Maps of Dust Infrared Emission for Use in Estimation of Reddening and Cosmic Microwave Background Radiation Foregrounds}",
      journal = {\apj},
         year = 1998,
        month = jun,
       volume = {500},
       number = {2},
        pages = {525-553},
          doi = {10.1086/305772},
archivePrefix = {arXiv},
       eprint = {astro-ph/9710327},
 primaryClass = {astro-ph},
       adsurl = {https://ui.adsabs.harvard.edu/abs/1998ApJ...500..525S}
}

@ARTICLE{Schlafly2011,
       author = {{Schlafly}, Edward F. and {Finkbeiner}, Douglas P.},
        title = "{Measuring Reddening with Sloan Digital Sky Survey Stellar Spectra and Recalibrating SFD}",
      journal = {\apj},
         year = 2011,
        month = aug,
       volume = {737},
       number = {2},
          eid = {103},
        pages = {103},
          doi = {10.1088/0004-637X/737/2/103},
archivePrefix = {arXiv},
       eprint = {1012.4804},
 primaryClass = {astro-ph.GA},
       adsurl = {https://ui.adsabs.harvard.edu/abs/2011ApJ...737..103S}
}

@PHDTHESIS{Stadel2001,
       author = {{Stadel}, Joachim Gerhard},
        title = "{Cosmological N-body simulations and their analysis}",
       school = {University of Washington, Seattle},
         year = 2001,
        month = jan,
       adsurl = {https://ui.adsabs.harvard.edu/abs/2001PhDT........21S}
}

@ARTICLE{Ruiz-Dern2018,
       author = {{Ruiz-Dern}, L. and {Babusiaux}, C. and {Arenou}, F. and {Turon}, C. and {Lallement}, R.},
        title = "{Empirical photometric calibration of the Gaia red clump: Colours, effective temperature, and absolute magnitude}",
      journal = {\aap},
         year = 2018,
        month = jan,
       volume = {609},
          eid = {A116},
        pages = {A116},
          doi = {10.1051/0004-6361/201731572},
archivePrefix = {arXiv},
       eprint = {1710.05803},
 primaryClass = {astro-ph.SR},
       adsurl = {https://ui.adsabs.harvard.edu/abs/2018A&A...609A.116R}
}

@ARTICLE{Haschke2012a,
       author = {{Haschke}, Raoul and {Grebel}, Eva K. and {Duffau}, Sonia},
        title = "{Three-dimensional Maps of the Magellanic Clouds using RR Lyrae Stars and Cepheids. I. The Large Magellanic Cloud}",
      journal = {\aj},
         year = 2012,
        month = oct,
       volume = {144},
       number = {4},
          eid = {106},
        pages = {106},
          doi = {10.1088/0004-6256/144/4/106},
archivePrefix = {arXiv},
       eprint = {1207.5791},
 primaryClass = {astro-ph.GA},
       adsurl = {https://ui.adsabs.harvard.edu/abs/2012AJ....144..106H}
}

@ARTICLE{Girardi1999,
       author = {{Girardi}, L{\'e}o},
        title = "{A secondary clump of red giant stars: why and where}",
      journal = {\mnras},
         year = 1999,
        month = sep,
       volume = {308},
       number = {3},
        pages = {818-832},
          doi = {10.1046/j.1365-8711.1999.02746.x},
archivePrefix = {arXiv},
       eprint = {astro-ph/9901319},
 primaryClass = {astro-ph},
       adsurl = {https://ui.adsabs.harvard.edu/abs/1999MNRAS.308..818G}
}

@ARTICLE{Girardi2016,
       author = {{Girardi}, L{\'e}o},
        title = "{Red Clump Stars}",
      journal = {\araa},
         year = 2016,
        month = sep,
       volume = {54},
        pages = {95-133},
          doi = {10.1146/annurev-astro-081915-023354},
       adsurl = {https://ui.adsabs.harvard.edu/abs/2016ARA&A..54...95G}
}

@ARTICLE{Lindegren2018,
       author = {{Lindegren}, L. and {Hern{\'a}ndez}, J. and {Bombrun}, A. and {Klioner}, S. and {Bastian}, U. and {Ramos-Lerate}, M. and {de Torres}, A. and {Steidelm{\"u}ller}, H. and {Stephenson}, C. and {Hobbs}, D. and {Lammers}, U. and {Biermann}, M. and {Geyer}, R. and {Hilger}, T. and {Michalik}, D. and {Stampa}, U. and {McMillan}, P.~J. and {Casta{\~n}eda}, J. and {Clotet}, M. and {Comoretto}, G. and {Davidson}, M. and {Fabricius}, C. and {Gracia}, G. and {Hambly}, N.~C. and {Hutton}, A. and {Mora}, A. and {Portell}, J. and {van Leeuwen}, F. and {Abbas}, U. and {Abreu}, A. and {Altmann}, M. and {Andrei}, A. and {Anglada}, E. and {Balaguer-N{\'u}{\~n}ez}, L. and {Barache}, C. and {Becciani}, U. and {Bertone}, S. and {Bianchi}, L. and {Bouquillon}, S. and {Bourda}, G. and {Br{\"u}semeister}, T. and {Bucciarelli}, B. and {Busonero}, D. and {Buzzi}, R. and {Cancelliere}, R. and {Carlucci}, T. and {Charlot}, P. and {Cheek}, N. and {Crosta}, M. and {Crowley}, C. and {de Bruijne}, J. and {de Felice}, F. and {Drimmel}, R. and {Esquej}, P. and {Fienga}, A. and {Fraile}, E. and {Gai}, M. and {Garralda}, N. and {Gonz{\'a}lez-Vidal}, J.~J. and {Guerra}, R. and {Hauser}, M. and {Hofmann}, W. and {Holl}, B. and {Jordan}, S. and {Lattanzi}, M.~G. and {Lenhardt}, H. and {Liao}, S. and {Licata}, E. and {Lister}, T. and {L{\"o}ffler}, W. and {Marchant}, J. and {Martin-Fleitas}, J.-M. and {Messineo}, R. and {Mignard}, F. and {Morbidelli}, R. and {Poggio}, E. and {Riva}, A. and {Rowell}, N. and {Salguero}, E. and {Sarasso}, M. and {Sciacca}, E. and {Siddiqui}, H. and {Smart}, R.~L. and {Spagna}, A. and {Steele}, I. and {Taris}, F. and {Torra}, J. and {van Elteren}, A. and {van Reeven}, W. and {Vecchiato}, A.},
        title = "{Gaia Data Release 2. The astrometric solution}",
      journal = {\aap},
         year = 2018,
        month = aug,
       volume = {616},
          eid = {A2},
        pages = {A2},
          doi = {10.1051/0004-6361/201832727},
archivePrefix = {arXiv},
       eprint = {1804.09366},
 primaryClass = {astro-ph.IM},
       adsurl = {https://ui.adsabs.harvard.edu/abs/2018A&A...616A...2L}
}

@ARTICLE{Fitzpatrick1999,
       author = {{Fitzpatrick}, Edward L.},
        title = "{Correcting for the Effects of Interstellar Extinction}",
      journal = {\pasp},
         year = 1999,
        month = jan,
       volume = {111},
       number = {755},
        pages = {63-75},
          doi = {10.1086/316293},
archivePrefix = {arXiv},
       eprint = {astro-ph/9809387},
 primaryClass = {astro-ph},
       adsurl = {https://ui.adsabs.harvard.edu/abs/1999PASP..111...63F}
}

@ARTICLE{Graczyk2020,
       author = {{Graczyk}, Dariusz and {Pietrzy{\'n}ski}, Grzegorz and {Thompson}, Ian B. and {Gieren}, Wolfgang and {Zgirski}, Bart{\l}omiej and {Villanova}, Sandro and {G{\'o}rski}, Marek and {Wielg{\'o}rski}, Piotr and {Karczmarek}, Paulina and {Narloch}, Weronika and {Pilecki}, Bogumi{\l} and {Taormina}, Monica and {Smolec}, Rados{\l}aw and {Suchomska}, Ksenia and {Gallenne}, Alexandre and {Nardetto}, Nicolas and {Storm}, Jesper and {Kudritzki}, Rolf-Peter and {Ka{\l}uszy{\'n}ski}, Miko{\l}aj and {Pych}, Wojciech},
        title = "{A Distance Determination to the Small Magellanic Cloud with an Accuracy of Better than Two Percent Based on Late-type Eclipsing Binary Stars}",
      journal = {\apj},
         year = 2020,
        month = nov,
       volume = {904},
       number = {1},
          eid = {13},
        pages = {13},
          doi = {10.3847/1538-4357/abbb2b},
archivePrefix = {arXiv},
       eprint = {2010.08754},
 primaryClass = {astro-ph.GA},
       adsurl = {https://ui.adsabs.harvard.edu/abs/2020ApJ...904...13G}
}

@ARTICLE{Nidever2008,
       author = {{Nidever}, David L. and {Majewski}, Steven R. and {Butler Burton}, W.},
        title = "{The Origin of the Magellanic Stream and Its Leading Arm}",
      journal = {\apj},
         year = 2008,
        month = may,
       volume = {679},
       number = {1},
        pages = {432-459},
          doi = {10.1086/587042},
       adsurl = {https://ui.adsabs.harvard.edu/abs/2008ApJ...679..432N}
}

\end{document}